\documentclass[fleqn,12pt]{article}
\usepackage[numbers,sort&compress]{natbib}
\usepackage[footnotesep=0.4in]{geometry}
\usepackage{graphicx}
\usepackage{amsmath}
\usepackage{bbding}
\usepackage{pifont}
\usepackage{amsfonts}
\usepackage{caption2}
\usepackage{float}
\usepackage{graphics}
\usepackage{subfigure}
\usepackage{float}
\usepackage{booktabs}
\usepackage{ctex}
\makeatletter
\newcommand\figcaption{\def\@captype{\textbf{figure}}\caption}
\newcommand\tabcaption{\def\@captype{table}\caption}
\begin{document}
\date{}
\title{Stationary nonlinear waves, superposition modes and modulational instability characteristics  in the AB system}
\author{Lei Wang$^{1}$$^,$\thanks{Corresponding
author:  50901924@ncepu.edu.cn},\,  Zi-Qi Wang$^{2}$,  Jian-Hui Zhang$^{2}$, Feng-Hua Qi$^{4}$, Min Li$^{1}$
\\{\em 1. Department of Mathematics and Physics, North China Electric}\\
{\em  Power University, Beijing 102206, P.\ R.\ China}
\\{\em 2. School of Energy Power and Mechanical Engneering,}\\ {\em North China Electric}
{\em  Power University, Beijing 102206, P.\ R.\ China}\\
{\em 3. School of Information, Beijing Wuzi University, }\\
{\em  Beijing 101149, P.\ R.\ China}
}
\maketitle

\newpage
\begin{abstract}
We study the AB system describing marginally unstable baroclinic wave packets in geophysical fluids and also ultra-short pulses in
nonlinear optics. We show that the breather can be converted into different types of stationary nonlinear waves on constant backgrounds,  including the multi-peak soliton, M-shaped soliton, W-shaped soliton and periodic wave. We also investigate the nonlinear interactions between these waves, which display some novel patterns due to the non-propagating characteristics of the  solitons: (1) Two antidark solitons can produce a W-shaped soliton instead of a higher-order antidark one; (2)  The interaction between an antidark soliton and a W-shaped soliton  can not only generate a higher-order antidark soliton, but also form a W-shaped solion pair; (3) The interactions between an oscillation W-shaped soliton and an oscillation M-shaped soliton show the multi-peak structures. We find that the transition occurs at a modulational stability region in a low perturbation frequency region.

\textbf{Solitons, breathers, and rogue waves  have been  observed   widely in nonlinear optics  and fluid mechanics. Recent studies have revealed the intricate relation between the soliton and breather (or rogue wave) solutions of certain higher-order and coupled nonlinear evolution equations. Breathers or rogue waves may be converted into various nonlinear waves on constant backgrounds. The interactions among these transformed waves show some novel characteristics.  Hereby, we will consider the  AB system describing marginally unstable baroclinic wave packets in geophysical fluids and also ultra-short pulses in nonlinear optics. Several types of transformed stationary nonlinear waves will be demonstrated under some special conditions. Due to the non-propagating characteristic, the nonlinear superpositions  include some  interesting phenomena. Further, the transformed nonlinear waves are associated with a  modulational stability region in a low perturbation frequency region.}\\

\end{abstract}
$\,$\vspace{2mm}
\newpage
\noindent\textbf{\Large{\uppercase\expandafter{1}. Introduction}}

Echoing soliton concepts that flourished in the multidisciplinary since a few decades ago, rogue wave has attracted recently the attention of researchers in various physical settings, e.g., in hydrodynamics~\cite{HY}, capillary waves~\cite{CW}, plasma physics~\cite{PP}, nonlinear optics~\cite{NO}, and Bose-Einstein condensation~\cite{BEC}.
Due to their harm to various hydrotechnic constructions, the investigation of rogue waves becomes a very important problem~\cite{DR}. Rogue waves, which have a peak amplitude generally more than twice the significant wave height, are thought to appear from nowhere and disappear without a trace in the ocean and most experimental optical systems~\cite{RWD}. Among various models describing such waves, the focusing nonlinear Schr\"{o}dinger  (NLS) equation~\cite{NLSE} is the most accepted one. The NLS equation admits a type of rational solution that is localized in both space and time, i.e., the Peregine soliton~\cite{PS}. This simplest rational solution of the NLS equation, which was first predicted as far as 30 years ago~\cite{PS}, is frequently used as a model of a rogue
wave~\cite{RWY}. Breather solutions are presently regarded as potential prototypes for the rogue waves in many fields of physics~\cite{BREA}. Breathers develop due to the instability of small amplitude perturbations that may grow in size to disastrous proportions~\cite{BREA1}. There are two types of breathers, namely, the Kuznetsov-Ma breathers (KMBs)~\cite{KMB} and Akhmediev breathers (ABs)~\cite{ABB}. The KMBs are periodic in space and localized in time while the ABs are periodic in time and localized in space~\cite{KMB,ABB}. Taking the period of both solutions to infinity leads to a first-order doubly localized Peregrine soliton.

The modulational instability (MI) is generally considered to be one of factors which may give rise to rogue-wave excitation~\cite{PS}. As a fundamental characteristic of many nonlinear dispersive systems, MI is related to the growth of periodic perturbations on an unstable continuous-wave background~\cite{MI1}. A rogue wave may be the result of MI, but not every type of MI results in rogue wave generation~\cite{MI2}. One of theoretical researches has revealed the close relationship between rogue waves and baseband MI, i.e., MI whose bandwidth includes components of arbitrarily low frequency~\cite{MI3,WL}. Another possible explanation presented by Zhao and Ling is that rogue wave comes from MI under the ``resonance" perturbation with continuous wave background~\cite{MI5}.

Recent studies have revealed the intricate link between the rogue waves (or breathers) and solitons of a certain class of nonlinear evolution equations. When the eigenvalues meet a certain locus on the complex plane, Akhmediev  \emph{et al.}  have discovered that the breather solutions of the
Hirota~\cite{ZH1} equation and fifth-order NLS~\cite{ZH2} equation can be converted into soliton solutions on a background, which does not exist in the standard NLS equation. He \emph{et al.} have reported that the rational solution of a mixed NLS equation can describe five soliton states, including a
paired bright-bright soliton, a single soliton, a paired bright-grey soliton, a paired bright-black soliton, and a rogue wave state~\cite{ZH3}. They have found that the state transition among these five states is induced by tuning the effects of self-steepening and self-phase modulation. Liu  \emph{et al.}  have shown that the breathers can be converted into different types of nonlinear waves in the coupled NLS-MB system, including the multipeak soliton, periodic wave, antidark soliton, and W-shaped soliton~\cite{ZH4}. In particular, they have indicated that the transition between the rogue
waves and W-shaped solitons of the Hirota and coupled Hirota equations occurs as a result of the attenuation of MI growth rate to vanishing in the zero-frequency perturbation region~\cite{ZH5}.

In this paper,  we study the AB system~\cite{AB,AB1},
\begin{equation}\label{AB1}
\begin{aligned}
&(\frac{\partial}{\partial T}+c_{1}\,\frac{\partial}{\partial X})(\frac{\partial}{\partial T}+c_{2}\,\frac{\partial}{\partial X})\,A=
l_{1}\,A-l_{2}\,A\,B,\\
&(\frac{\partial}{\partial T}+c_{2}\,\frac{\partial}{\partial X})\,B=(\frac{\partial}{\partial T}+c_{1}\,\frac{\partial}{\partial X})|A|^{2}\,,
\end{aligned}
\end{equation}
where $A$ is the amplitude of the wave packet and $ B$ is a quantity measuring the correction of the basic flow due to the wave packet, $T$ and $X$ denote the time and space variables, respectively, $c_{1}$
and $c_{2}$ stand for two group velocities of the underlying linear problem, and $l_{1}$ represents a parameter measuring the state of the basic
flow. When the basic flow is super-critical, $ l_{1} > 0$, and when the basic
flow is sub-critical, $l_{1} < 0$. The parameter $l_{2}$ reflects the interaction of the wave packet and the meanflow, and it is always positive. System~(\ref{AB1}) describes marginally unstable baroclinic wave packets in geophysical fluids and also ultra-short pulses in nonlinear
optics.

As shown in Ref.~\cite{CAB}, System~(\ref{AB1}) can be reduced to a simpler form
\begin{equation}\label{AB}
A_{xt}=\alpha\,A+\beta\,A\,B,\quad
B_{x}=-\frac{1}{2}\gamma(|A|^{2})_{t}\,,
\end{equation}
with
\begin{equation}\label{AB11}
\alpha=\frac{n_{1}\,c_{2}}{(c_{1}-c_{2})^{2}}\,,\quad
\beta=-\frac{n_{2}\,c_{2}}{(c_{1}-c_{2})^{2}}\,,\quad
\gamma=-\frac{2}{c_{2}}\,,
\end{equation}
by the transformations
\begin{equation}\label{AB111}
x=X-c_{1}\,T ,\quad t=T-\frac{X}{c_{2}}\,.
\end{equation}
Recently, certain properties of System~(\ref{AB}) have been investigated. System~(\ref{AB}) can be transformed to the Sine-Gordon equation when $A$ is the real value and to the self-induced transparency system when $A$ is the complex value~\cite{AB,AB1}. Lax pair and some periodic solutions of System~(\ref{AB}) have been derived in Ref.~\cite{AB1}.  MI and breather dynamics of System~(\ref{AB}) have been discussed in Ref.~\cite{GR}. Ref.~\cite{TBK} has studied the envelope solitary waves and periodic waves of System~(\ref{AB}). The higher-order rogue wave solutions of System~(\ref{AB}) have been found via the modified Darboux transformation (mDT) in Ref.~\cite{WX}. The semirational solutions have been
derived in Ref.~\cite{WL} and the link between the baseband MI and the existence condition of these rogue waves has been revealed. Recently, the rogue waves and MI have been demonstrated for a coupled AB system, i.e., a wave-current interaction model describing baroclinic instability processes in geophysical flows~\cite{CAB}.

The aim of the present paper is to study the breather-soliton dynamics of System~(\ref{AB}). We present intriguing different kinds of nonlinear localized and periodic waves, including the multi-peak soliton, M-shaped soliton, W-shaped soliton and periodic wave. Interestingly, these waves are stationary nonlinear waves with respect to $x$-axis. Further, due to the nonpropagating  characteristic,  the nonlinear interactions show some novel properties. The breather-to-soliton transition is related to a special type of MI  analysis that involves a MI  stability region in a low perturbation frequency region.

The arrangement of the paper is as follows: In Sec.~2, we will present different types of stationary nonlinear waves of  System~(\ref{AB}),  including the multi-peak soliton, M-shaped soliton, W-shaped soliton and periodic waves. And the transition condition will be analytically given.  The properties of interactions between different types of nonlinear waves will be graphically studied in Sec.~3.  The connection between the MI growth rate and transition condition will be revealed in Sec.~4. Finally, Sec.~5 will be the conclusions of this paper.

\vspace{5mm}
\noindent\textbf{\Large{\uppercase\expandafter{2}. Different types of stationary nonlinear waves  }}

In this section, we mainly study the transitions between the first-order breather and various nonlinear waves  for System~(\ref{AB}). The first-order breather solution of System~(\ref{AB}) is given~\cite{WL}
\begin{equation}\label{EXP-B1}
\begin{aligned}
&A_{B}^{[1]}=\bigg(a+\,n_{1}\frac{G_{B}^{[1]}+i\,H_{B}^{[1]}}{D_{B}^{[1]}}\bigg)
\,e^{i\,\rho}\,,\\
&B_{B}^{[1]}=b+\frac{4\,i}{\beta}\,\bigg(\frac{m_{1}\,E_{B}^{[1]}+i\,n_{1}\,F_{B}^{[1]}}{D_{B}^{[1]}}\bigg)_{t}\,,
\end{aligned}
\end{equation}
 with
 \begin{equation}
\begin{aligned}
 &\rho=\omega\,x+k\,t\,,\qquad k=-\frac{\alpha+b\,\beta}{\omega}\,,\nonumber \\
 &G_{B}^{[1]}=k_{1}\,k_{2}\,\cos(t\,V_{H}+x\,h_{R})\cosh(2\,\chi_{I})-\cosh(t\,V_{T}+x\,h_{I})\sin(2\,\chi_{R})\,,\nonumber \\
 &H_{B}^{[1]}=k_{1}\,k_{2}\,\sin(t\,V_{H}+x\,h_{R})\sinh(2\,\chi_{I})+\cos(2\,\chi_{R})\sinh(t\,V_{T}+x\,h_{I})\,,\nonumber\\
 &D_{B}^{[1]}=-k_{1}\,k_{2}\,\cos(t\,V_{H}+x\,h_{R})\sin(2\,\chi_{R})+\cosh(t\,V_{T}+x\,h_{I})\cosh(2\,\chi_{I})\,,\nonumber\\
 &E_{B}^{[1]}=k_{1}\,k_{2}\,\cos(t\,V_{H}+x\,h_{R})\sin(2\,\chi_{R})-\cosh(t\,V_{T}+x\,h_{I})\cosh(2\,\chi_{I})\,,\nonumber\\
 &F_{B}^{[1]}=k_{1}\,k_{2}\,\sin(t\,V_{H}+x\,h_{R})\cos(2\,\chi_{R})-\sinh(t\,V_{T}+x\,h_{I})\sinh(2\,\chi_{I})\,,\nonumber\\
 &h=\sqrt{a^{2}\beta\gamma+(2\,\lambda+\omega)^{2}}=h_{R}+i\,h_{I}\,,\nonumber \\
 &\chi=\frac{1}{2}\,\arccos(\frac{h}{2})\,,\quad\,\varpi=\frac{h}{2}\,\Big(x+\frac{k}{2\,\lambda}\,t\Big)=(x+(\varpi_R+i\, \varpi_I)t)\frac{h}{2}\,,\notag \nonumber \\
 &V_{T}=2(\varpi_{R}\,h_{I}+\varpi_{I}\,h_{R})\,,\qquad V_{H}=2(\varpi_{R}\,h_{R}-\varpi_{I}\,h_{I})\,,\nonumber\\
 &k_{1}=1,\,k_{2}=\pm1 \,.\nonumber
\end{aligned}
 \end{equation}
From the above expression, one can find that the breather solution~(\ref{EXP-B1}) comprises the hyperbolic functions $\sinh F$ ($\cosh F$) and the trigonometric functions $\sin G$ ($\cos G$), where $\varpi_{R}+\frac{\varpi_{I}h_{R}}{h_{I}}$ and $\varpi_{R}-\frac{\varpi_{I}h_{I}}{h_{R}}$ are the corresponding velocities. The hyperbolic functions and trigonometric functions, respectively, characterize the localization and the periodicity of the transverse distribution $t$ of those waves. The
nonlinear wave described by the solution~(\ref{EXP-B1}) can be deemed to a nonlinear combination of a soliton and a periodic wave with the
velocities $\varpi_{R}+\frac{\varpi_{I}h_{R}}{h_{I}}$ and $\varpi_{R}-\frac{\varpi_{I}h_{I}}{h_{R}}$. In the following, we show that the solution~(\ref{EXP-B1}) can describe different kinds of nonlinear wave states depending on the values of velocity difference $\frac{\varpi_{I}(h_{R}^{2}+h_{I}^{2})}{h_{R}h_{I}}$.

When  $\varpi_{I}(\frac{h_{R}^{2}+h_{I}^{2}}{h_{R}h_{I}})\neq0$ (or $\varpi_{I}\neq0$), the solution~(\ref{EXP-B1}) characterizes the localized waves with breathing behavior on constant backgrounds (i.e., the breathers and rogue waves).  Further, if $m=-\frac{\omega}{2}$, we have the ABs with $|n|<|\frac{a\sqrt{\beta\gamma}}{2}|$, the KMBs with $|n|>|\frac{a\sqrt{\beta\gamma}}{2}|$, and the Peregrine soliton with $|n|=|\frac{a\sqrt{\beta\gamma}}{2}|$. Those solutions have been obtained in Refs.~\cite{WL,GR,WX}.

Conversely, if  $\varpi_{I}=0$, the soliton and a periodic wave in the solution~(\ref{EXP-B1}) have the same velocity $\varpi_{R}$. Meanwhile, we should point out that the case $\varpi_{I}=0$ is equivalent to the following condition
 \begin{equation}\label{KZFC1}
    \begin{aligned}
&\frac{V_{T}}{h_{I}}=\frac{V_{H}}{h_{R}}\,,
  \end{aligned}
 \end{equation}
 i.e.,
\begin{equation}\label{KZFC}
b=-\frac{\alpha}{\beta}\,.
\end{equation}
Eq.~(\ref{KZFC1}) means the extrema of trigonometric and hyperbolic functions in the solution~(\ref{EXP-B1}) is located along the same straight lines in the $(x,t)$-plane, which results in the transformation of the breather into a continuous soliton. Additionally, the case $b=-\frac{\alpha}{\beta}$ is also equivalent to $k=0$.
Then,  the parameters $\varpi_R$, $\varpi_I$, $V_T$ and $V_H$ vanish, and the solution~(\ref{EXP-B1}) only depends on $x$  and is independent of $t$. Thus, the nonlinear waves described by the solution~(\ref{EXP-B1}) possess the nonpropagating characteristic.

Under the transition condition~(\ref{KZFC}), we demonstrate several kinds of transformed nonlinear waves on constant background for $A$, including the multi-peak solitons, M-shaped soliton, periodic waves and W-shaped soliton. We omit the results in the $B$ component, since it only describes the plane wave in the case of $k=0$.
\textbf{Fig.~1} shows a multi-peak  soliton on constant background that does not propagate along $x$ direction.  Such structure is composed of a soliton and periodic wave. Adding the absolute value of real part to the eigenvalue,  $|m|$, we find that the peak numbers of multi-peak localized structure decrease, as depicted in \textbf{Fig.~2}. From the cross-sectional views, it is observed that both of those multi-peak solitons have one main peak which is located at $(0, 0)$. The maximum amplitude of $|A|^{2}$ at $(0, 0)$ can be given analytically
\begin{equation}\label{AP}
|A(0,0)|^{2}=(a-4\,\frac{n}{\sqrt{\beta\gamma}})^{2}\,.
\end{equation}
Eq.~(\ref{AP}) indicates that the imaginary part of eigenvalue ($n$) has real
effects on the maximum amplitude of $|A|^{2}$ at $(0, 0)$.

Further, the structure in Fig.~2 (we call it an oscillation W-shaped soliton) corresponds to the case $|A(0, 0)|^{2}_{xx}>0$ with the appropriate value of $n$ ($n=-0.7$). In other words, the coordinate origin $(0, 0)$ is a local maximum of $|A(x, 0)|^{2}$. Nevertheless, when the value of $n$ exceeds a certain range, $|A(0, 0)|^{2}_{xx}$ is less than zero, which means that the coordinate origin $(0, 0)$ is a local minimum of $|A(x, 0)|^{2}$. In this case, the oscillation W-shaped soliton  translates into a M-shaped soliton shown in \textbf{Fig.~3}. It is observed that this structure has two main peaks with identical amplitudes. In order to reveal the effect of the value of $n$ on $|A(0, 0)|^{2}_{xx}$, we plot \textbf{Fig.~4} with $a=2,\, b=1,\, \gamma=1,\,\omega=1,\,\alpha=-1,\,\beta=1,\,k_{1}=1,\,k_{2}=-1\,$ and $\, m=-0.8$. If $0<n<1.25$,  namely, $|A(0, 0)|^{2}_{xx}<0$, the solution~(\ref{EXP-B1}) describes the M-shaped soliton whereas it describes the W-shaped one.

\begin{figure}[H]\centering
\subfigure[]{\includegraphics[width=150 bp]{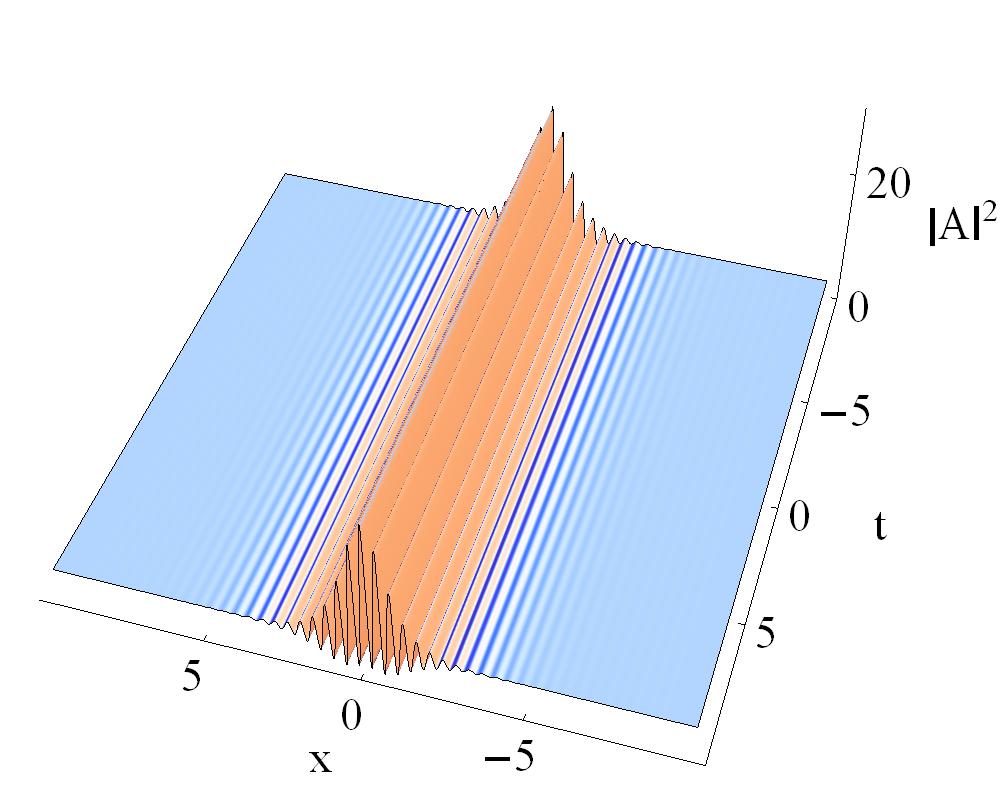}}
\quad
\subfigure[]{\includegraphics[width=100 bp]{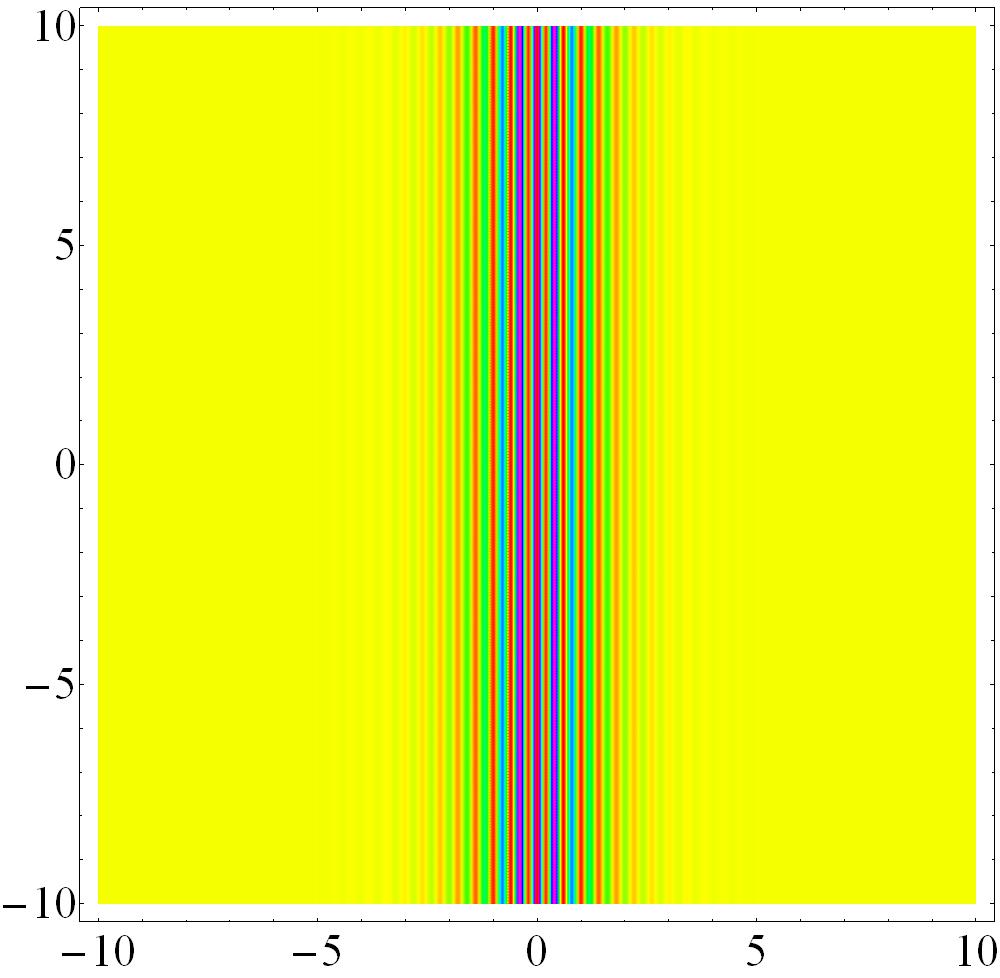}
\includegraphics[height=100 bp]{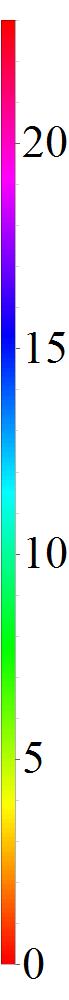}}
\quad
\subfigure[]{\includegraphics[width=130 bp]{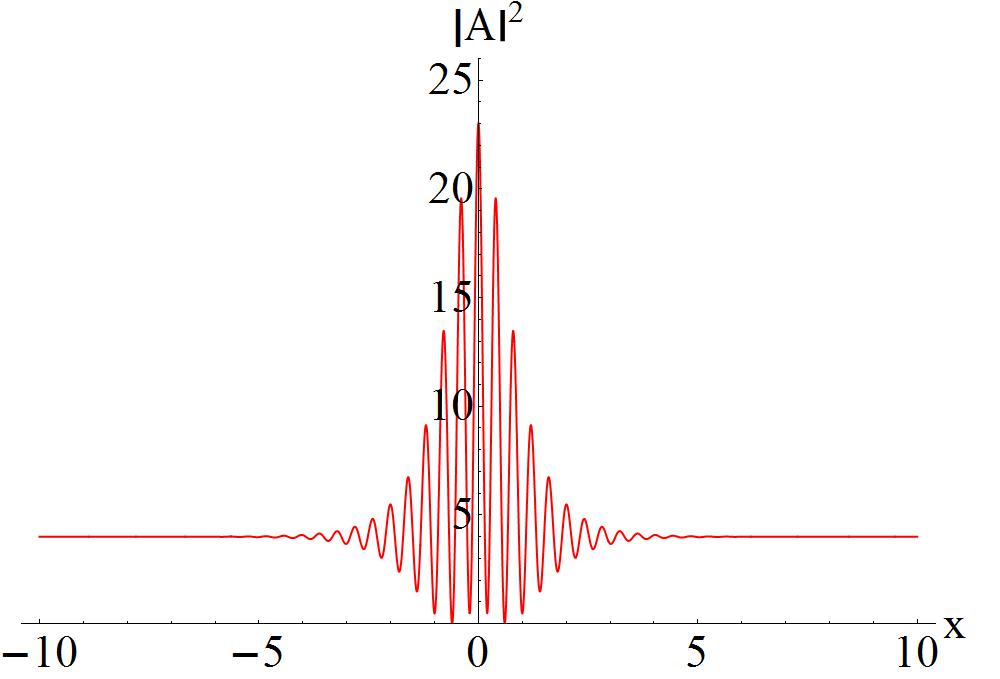}}
\caption{\footnotesize A breather transformed into a multi-peak  soliton with $a=2,\, b=1,\, \gamma=1,\,\omega=1,\,\alpha=-1,\,\beta=1,\,k_{1}=1,\,k_{2}=-1\,$ and $\, \lambda_{1}=\lambda_{2}^{*}=7.2-0.7\,i.$  (b) is the contour plot of (a). (c) is the  cross-sectional view  of (a) at $t=0$. }
\end{figure}
\begin{figure}[H]\centering
\subfigure[]{\includegraphics[width=150 bp]{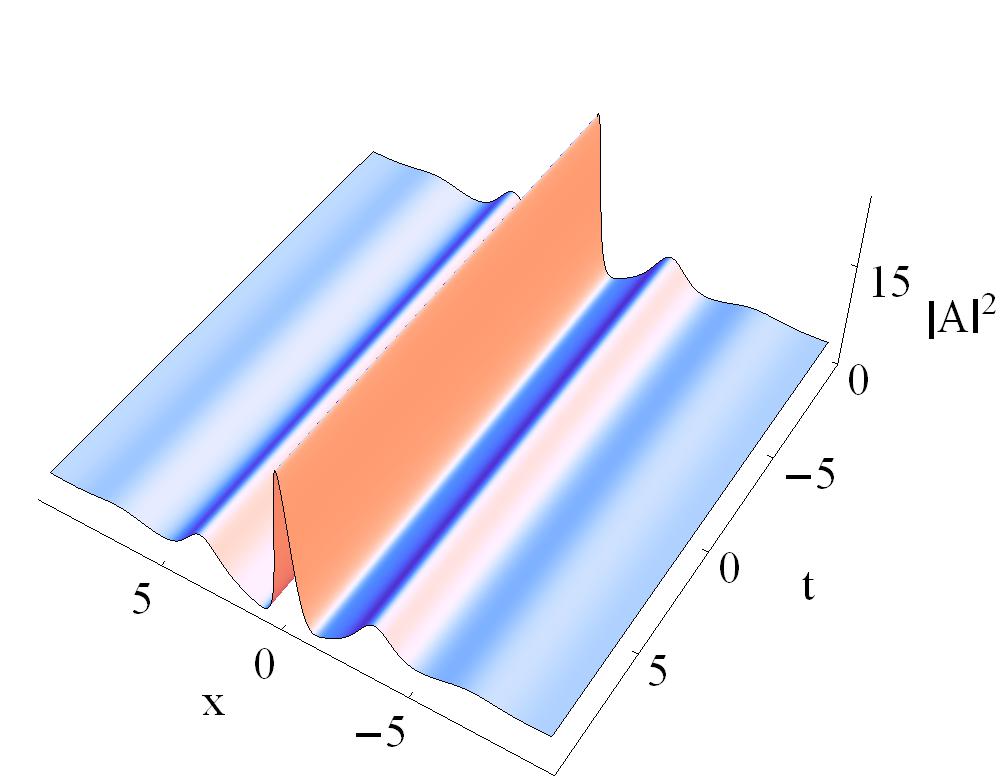}}
\quad
\subfigure[]{\includegraphics[width=100 bp]{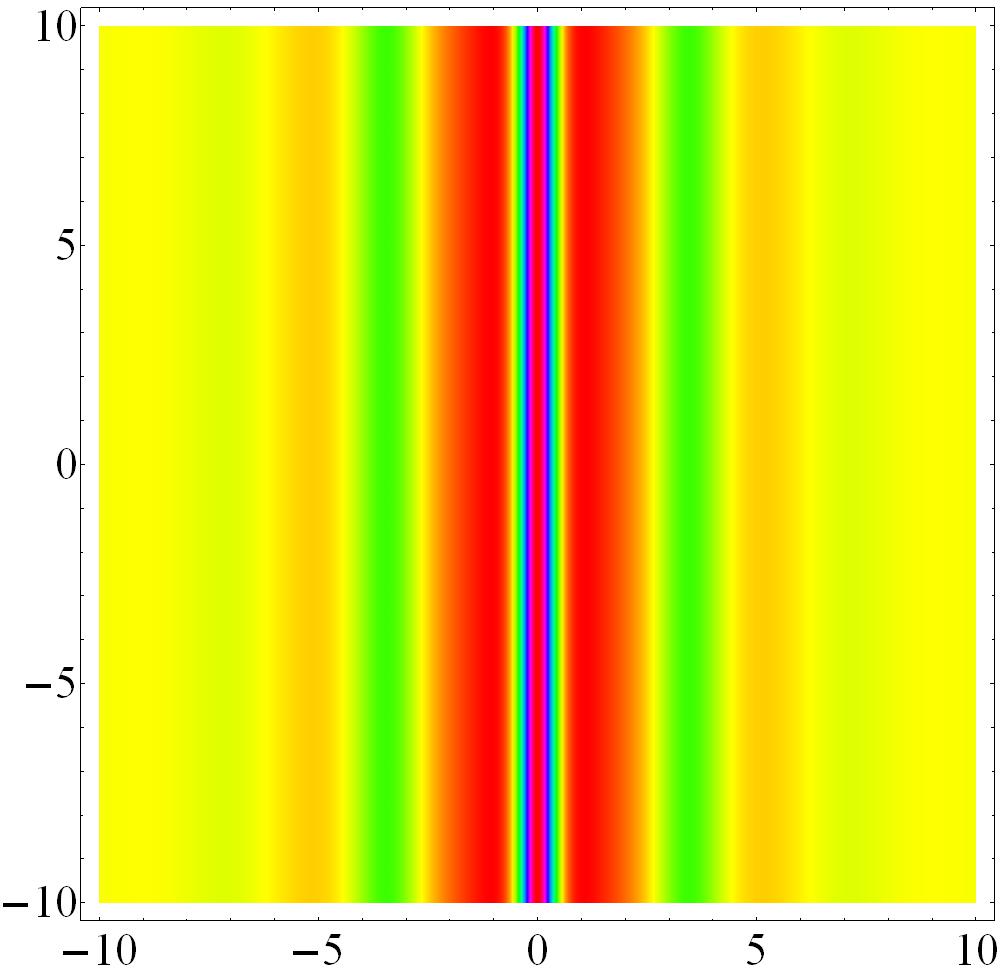}
\includegraphics[height=100 bp]{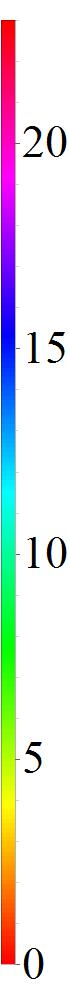}}
\quad
\subfigure[]{\includegraphics[width=130 bp]{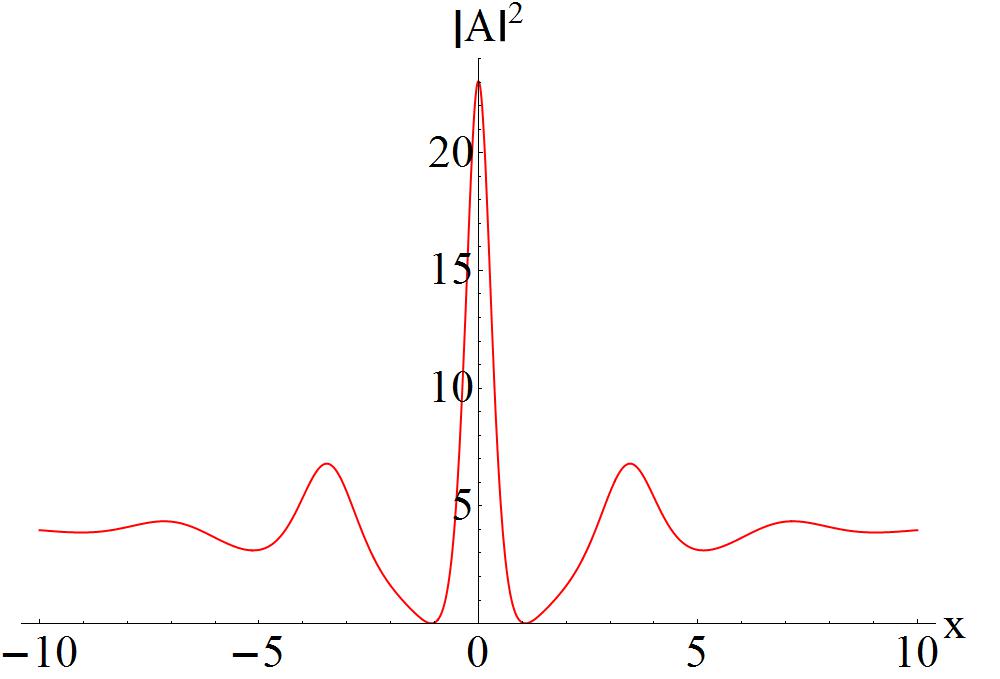}}
\caption{\footnotesize A breather transformed into an oscillation W-shaped soliton with \,$a=2,\, b=1,\, \gamma=1,\,\omega=1,\,\alpha=-1,\,\beta=1,\,k_{1}=1,\,k_{2}=-1\,$ and $\, \lambda_{1}=\lambda_{2}^{*}=-0.8-0.7\,i.$  (b) is the contour plot of (a). (c) is the  cross-sectional view  of (a) at $t=0$. }
\end{figure}

\begin{figure}[H]\centering
\subfigure[]{\includegraphics[width=150 bp]{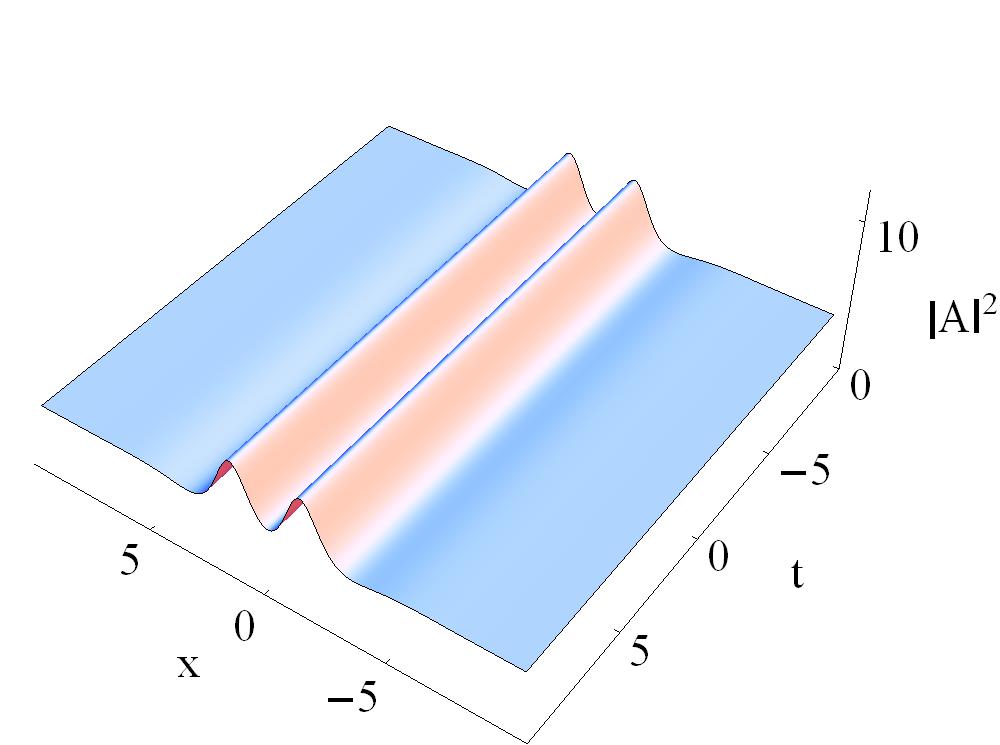}}
\quad
\subfigure[]{\includegraphics[width=100 bp]{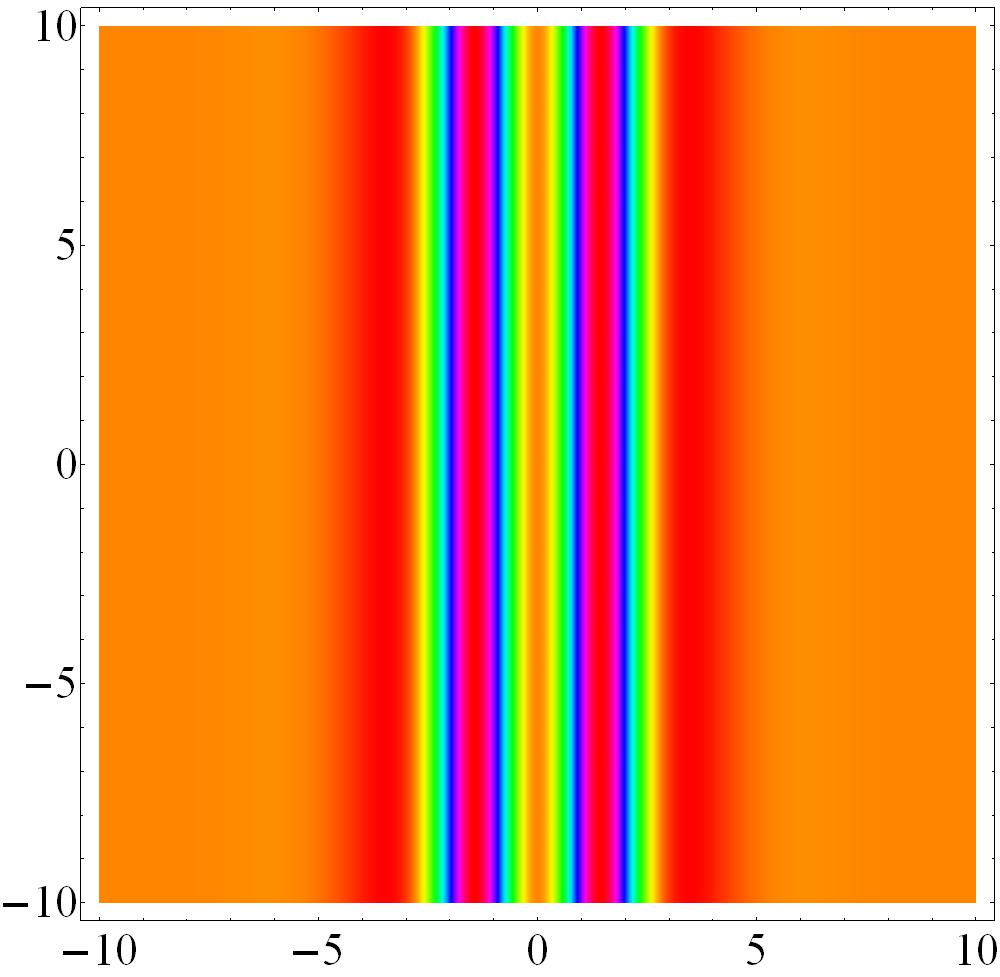}
\includegraphics[height=100 bp]{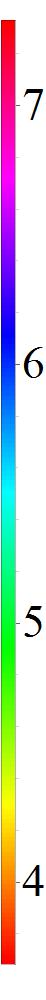}}
\quad
\subfigure[]{\includegraphics[width=130 bp]{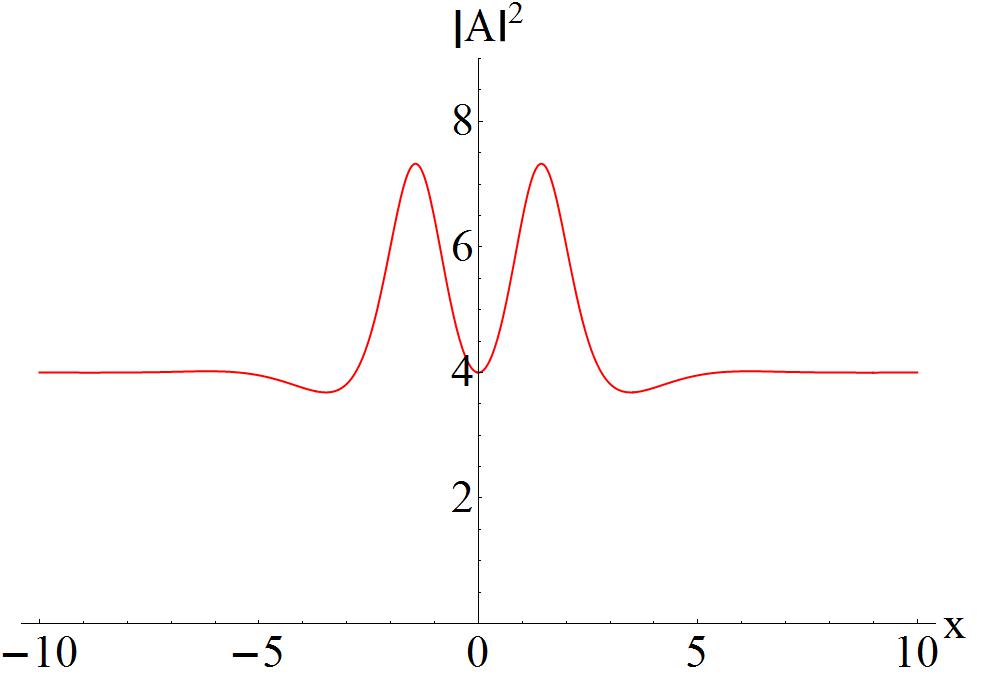}}
\caption{\footnotesize A breather transformed into a M-shaped soliton with
\,$a=2,\, b=1,\, \gamma=1,\,\omega=1,\,\alpha=-1,\,\beta=1,\,k_{1}=1,\,k_{2}=-1\,$ and $\, \lambda_{1}=\lambda_{2}^{*}=-0.8+\,i.$  (b) is the contour plot of (a). (c) is the  cross-sectional view  of (a) at $t=0$. }
\end{figure}

\begin{figure}[H]\centering
\quad
\includegraphics[width=180 bp]{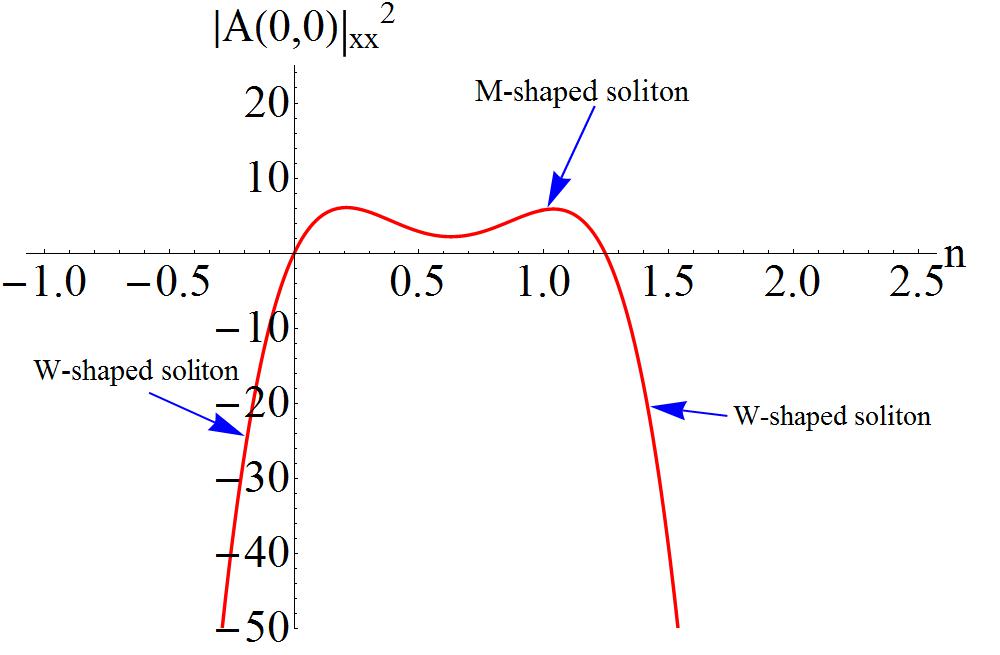}
\caption{\footnotesize Effects of the imaginary part of eigenvalue $n$ on $|A(0, 0)|^{2}_{xx}$  with
\,$a=2,\, b=1,\, \gamma=1,\,\omega=1,\,\alpha=-1,\,\beta=1,\,k_{1}=1,\,k_{2}=-1\,$ and $\, m=-0.8$. Two zeros of the $|A(0, 0)|^{2}_{xx}$ are in $(0, 0)$ and $(1.25, 0)$ respectively.}
\end{figure}

Next, we derive two special nonlinear waves from the solution~(\ref{EXP-B1}), i.e., the antidark soliton and periodic wave.
The former exists in isolation when $h_{R}$ vanishes, while the latter independently exists when $h_{I}$ vanishes. Therefore, the antidark soliton and periodic wave are shown in forms of exponential and trigonometric functions respectively. Specifically, the  analytical expressions read as, for the soliton,
\begin{equation}
\qquad\qquad\qquad  A_{S}^{[1]}=(a+\,n_{1}\frac{G_{S}^{[1]}+i\,H_{S}^{[1]}}{D_{S}^{[1]}})\,e^{i\,\rho}\,,\quad B_{S}^{[1]}=b\,,
\end{equation}
with
\begin{equation}
\begin{aligned}
&G_{S}^{[1]}=k_{1}\,k_{2}\,\cosh(2\,\chi_{I})-\cosh(x\,h_{I})\sin(2\,\chi_{R})\,, \\
&H_{S}^{[1]}=\cos(2\,\chi_{R})\sinh(x\,h_{I}),\\
&D_{S}^{[1]}=-k_{1}\,k_{2}\,\sin(2\,\chi_{R})+\cosh(x\,h_{I})\cosh(2\,\chi_{I})\,,\\
\end{aligned}
\end{equation}
and for the periodic wave,
\begin{equation}
\qquad\qquad\qquad A_{P}^{[1]}=(a+\,n_{1}\frac{G_{P}^{[1]}+i\,H_{P}^{[1]}}{D_{P}^{[1]}})\,e^{i\,\rho}\,,\quad B_{P}^{[1]}=b\,.
\end{equation}
with
\begin{equation}
\begin{aligned}
&G_{P}^{[1]}=k_{1}\,k_{2}\,\cos(x\,h_{R})\cosh(2\,\chi_{I})-\sin(2\,\chi_{R})\,, \\
&H_{P}^{[1]}=k_{1}\,k_{2}\,\sin(x\,h_{R})\sinh(2\,\chi_{I})\,,
\\
&D_{P}^{[1]}=-k_{1}\,k_{2}\,\cos(x\,h_{R})\sin(2\,\chi_{R})+\cosh(2\,\chi_{I})\,.
\\
\end{aligned}
\end{equation}
\begin{figure}[H]\centering
\subfigure[]{\includegraphics[width=150 bp]{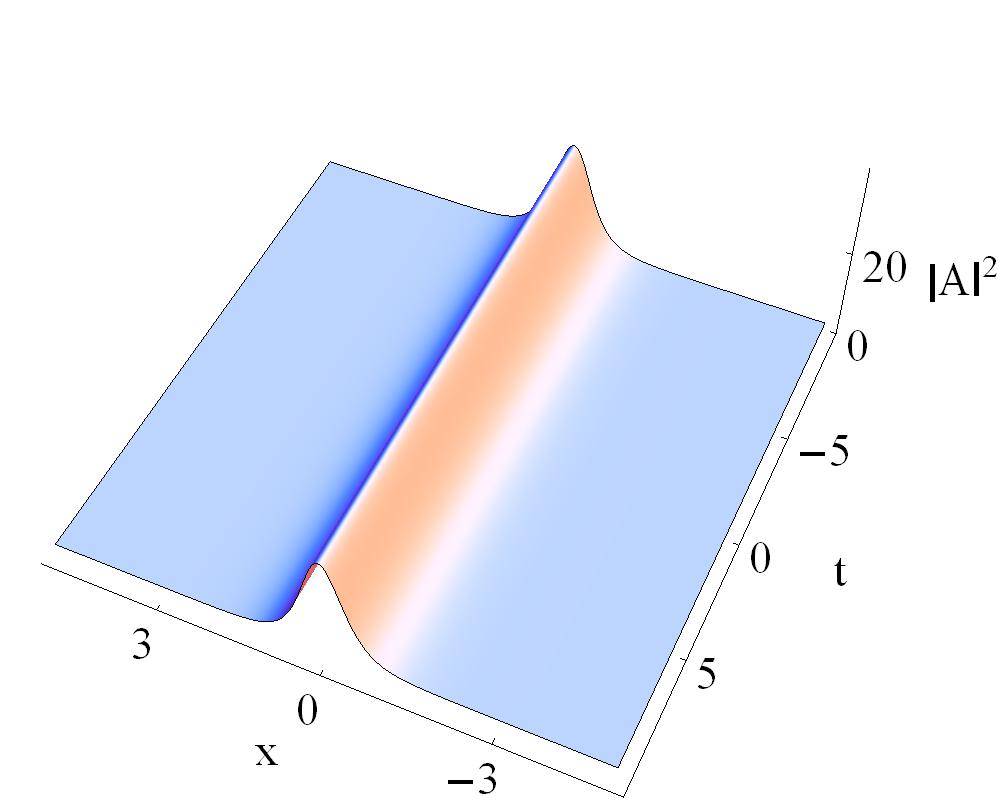}}
\quad
\subfigure[]{\includegraphics[width=100 bp]{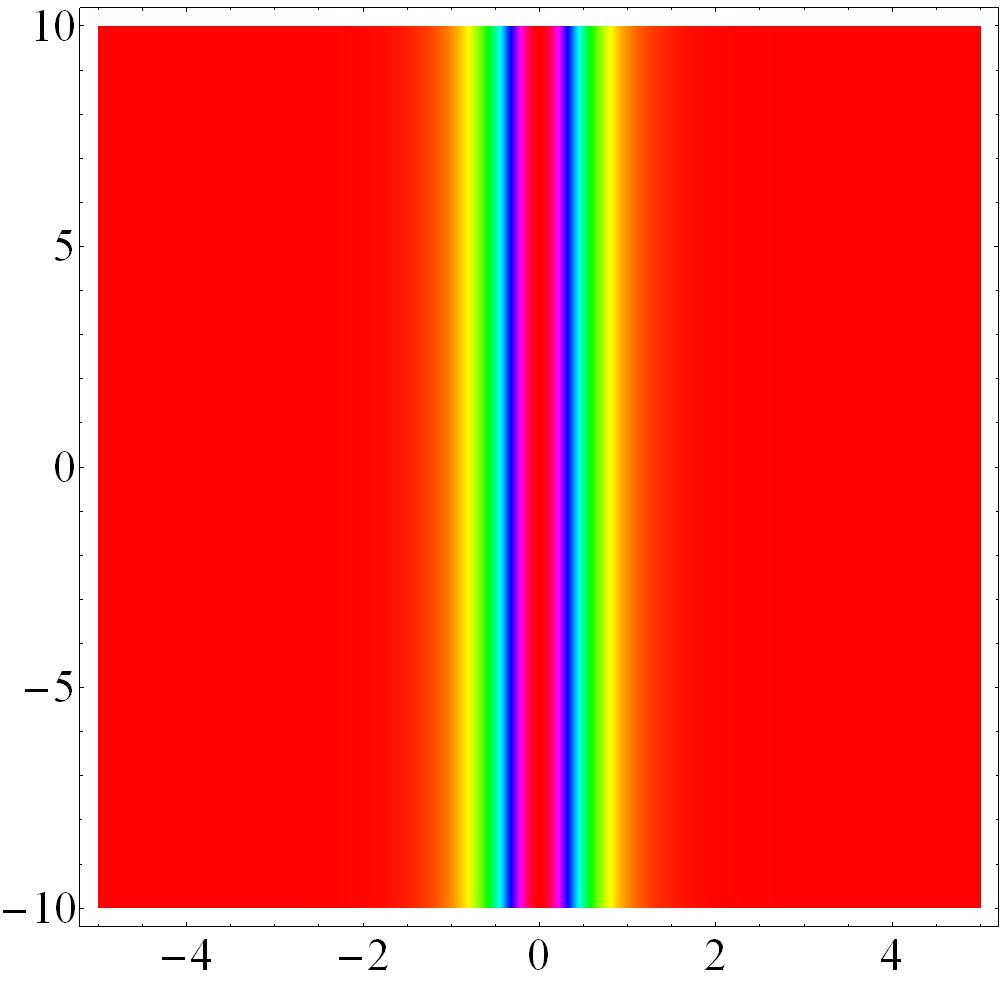}
\includegraphics[height=100 bp]{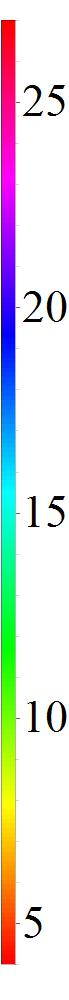}}
\quad
\subfigure[]{\includegraphics[width=130 bp]{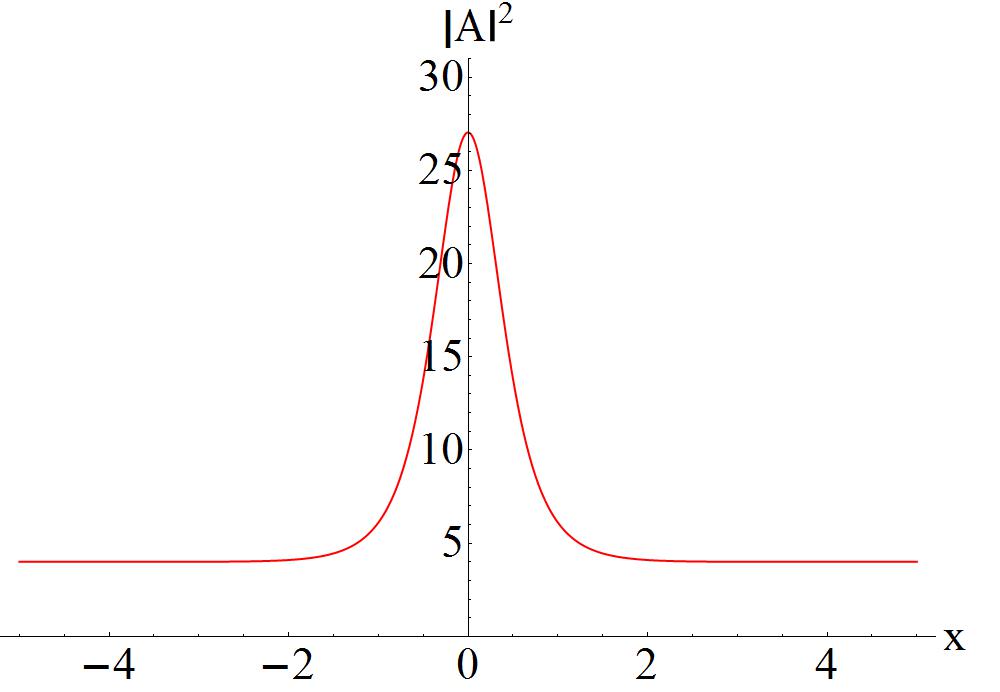}}
\caption{\footnotesize A breather transformed into an antidark  soliton with
\,$a=2,\, b=1,\, \gamma=1,\,\omega=1,\,\alpha=-1,\,\beta=1,\,k_{1}=1,\,k_{2}=1\,$ and $\, \lambda_{1}=\lambda_{2}^{*}=-0.5-1.8\,i.$ (b) is the contour plot of (a). (c) is the  cross-sectional view  of (a) at $t=0$. }
\end{figure}
\begin{figure}[H]\centering
\subfigure[]{\includegraphics[width=170 bp]{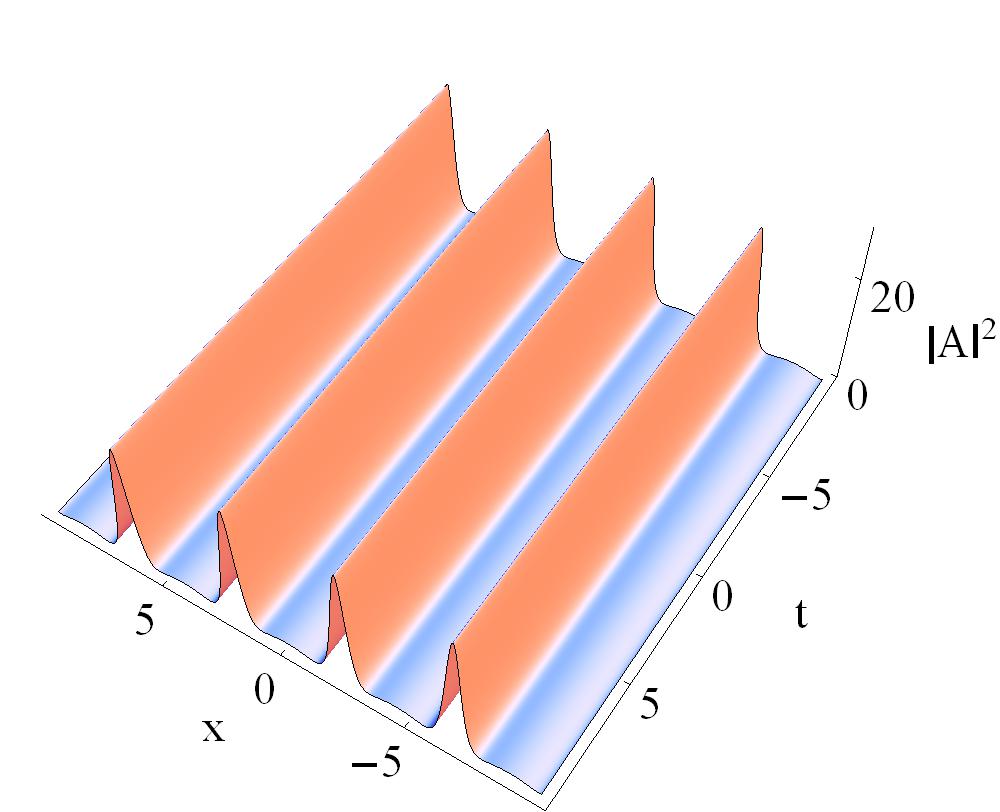}}
\quad
\subfigure[]{\includegraphics[width=100 bp]{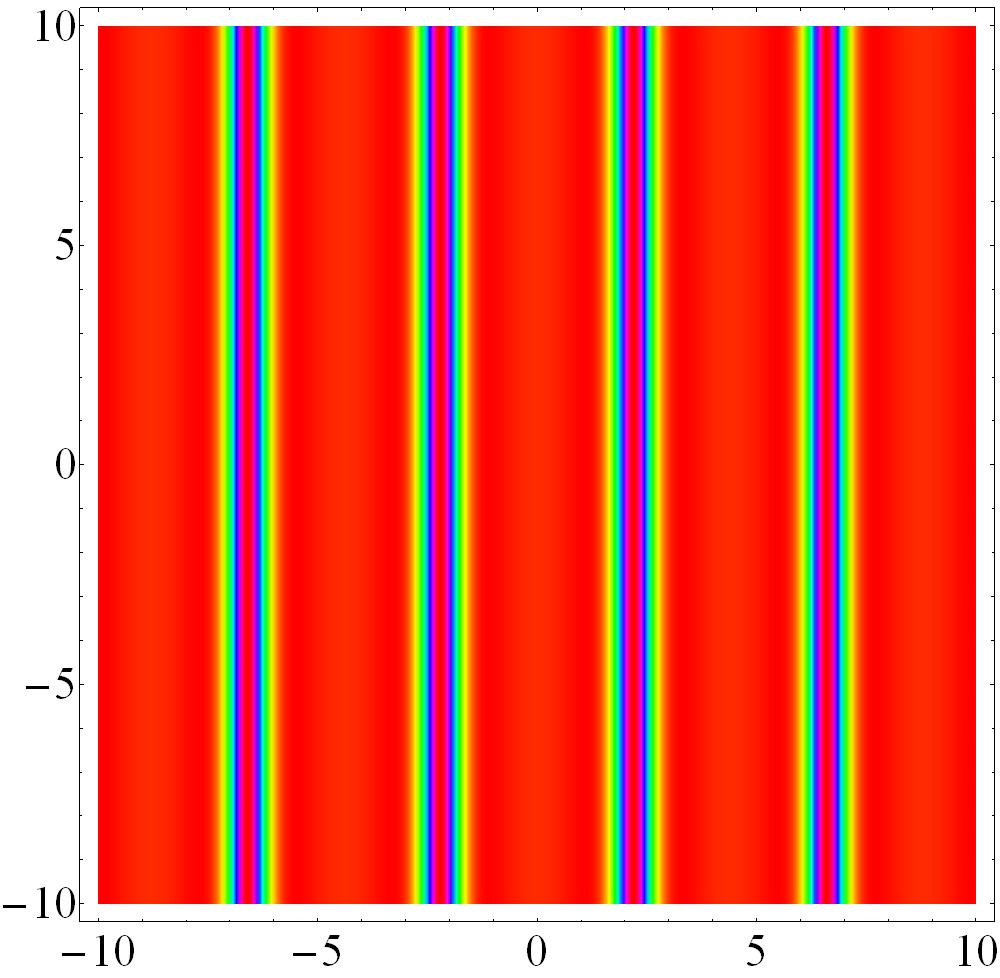}
\includegraphics[height=100 bp]{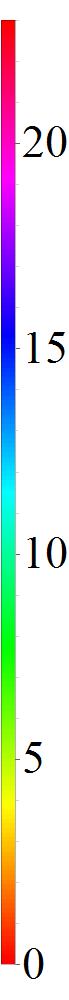}}
\quad
\subfigure[]{\includegraphics[width=130 bp]{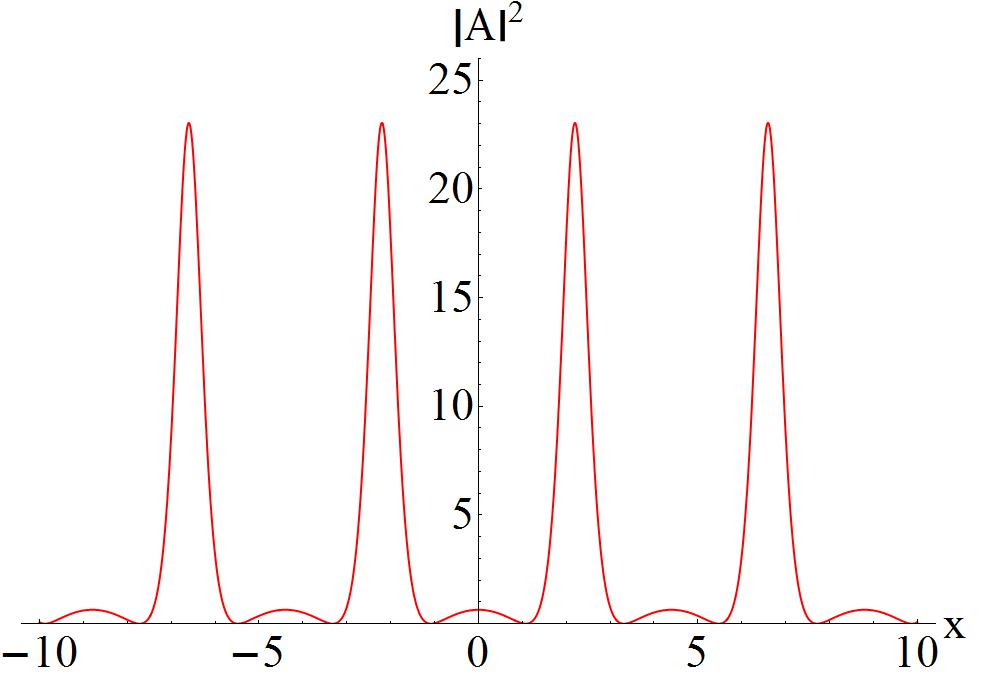}}
\caption{\footnotesize A breather transformed into a periodic wave with
\,$a=2,\, b=1,\, \gamma=1,\,\omega=1,\,\alpha=-1,\,\beta=1,\,k_{1}=1,\,k_{2}=-1\,$ and $\, \lambda_{1}=\lambda_{2}^{*}=-0.5+0.7\,i.$  (b) is the contour plot of (a). (c) is the  cross-sectional view  of (a) at $t=0$. }
\end{figure}

\textbf{Fig.~5} describes a soliton that does not propagate along $x$ direction. It is shown that this soliton lies on a plane-wave background with the peak $(a\sqrt{\beta\gamma}-2\,n)^{2}$. This kind of wave is referred to the antidark soliton which was firstly reported in the scalar NLS system with the third-order dispersion~\cite{ANTI}. Recent studies on the NLS-MB system have also presented the similar structures~\cite{ZH4}. Further, for the value of $a$ becomes zero, this soliton will be converted into a standard bright soliton. \textbf{Fig.~6} exhibits the periodic wave with the period $P=\frac{\pi}{h_{R}}$.  It is interesting, as it looks like higher-order wave but appears from the same solution.
\begin{figure}[H]
\centering
\subfigure[]{\includegraphics[width=170 bp]{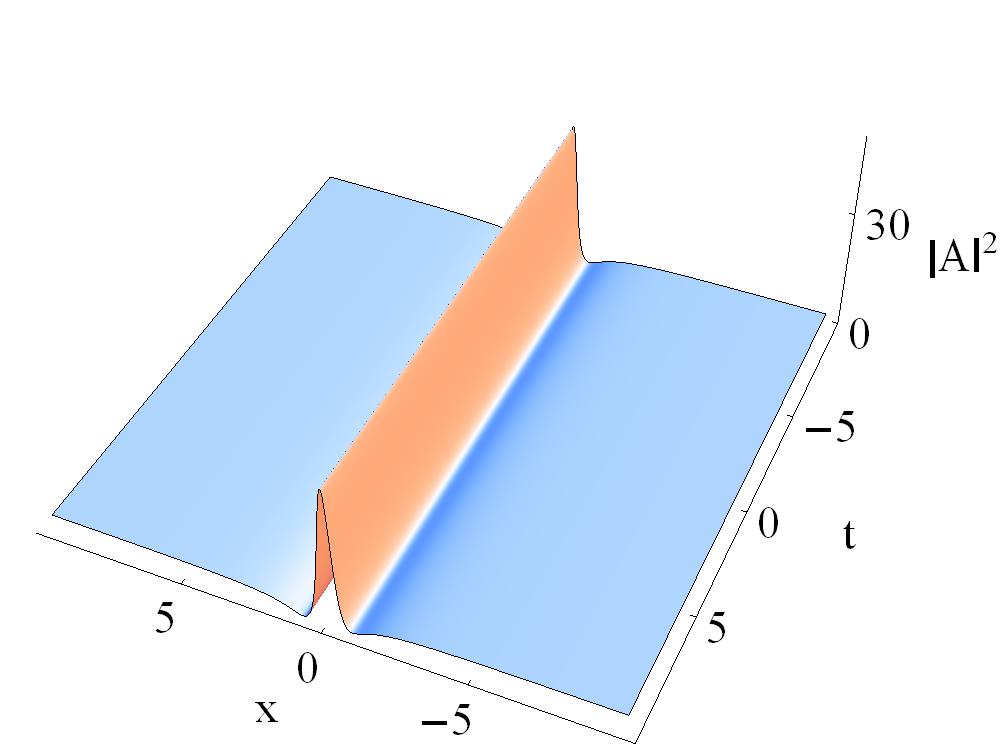}}
\quad
\subfigure[]{\includegraphics[width=90 bp]{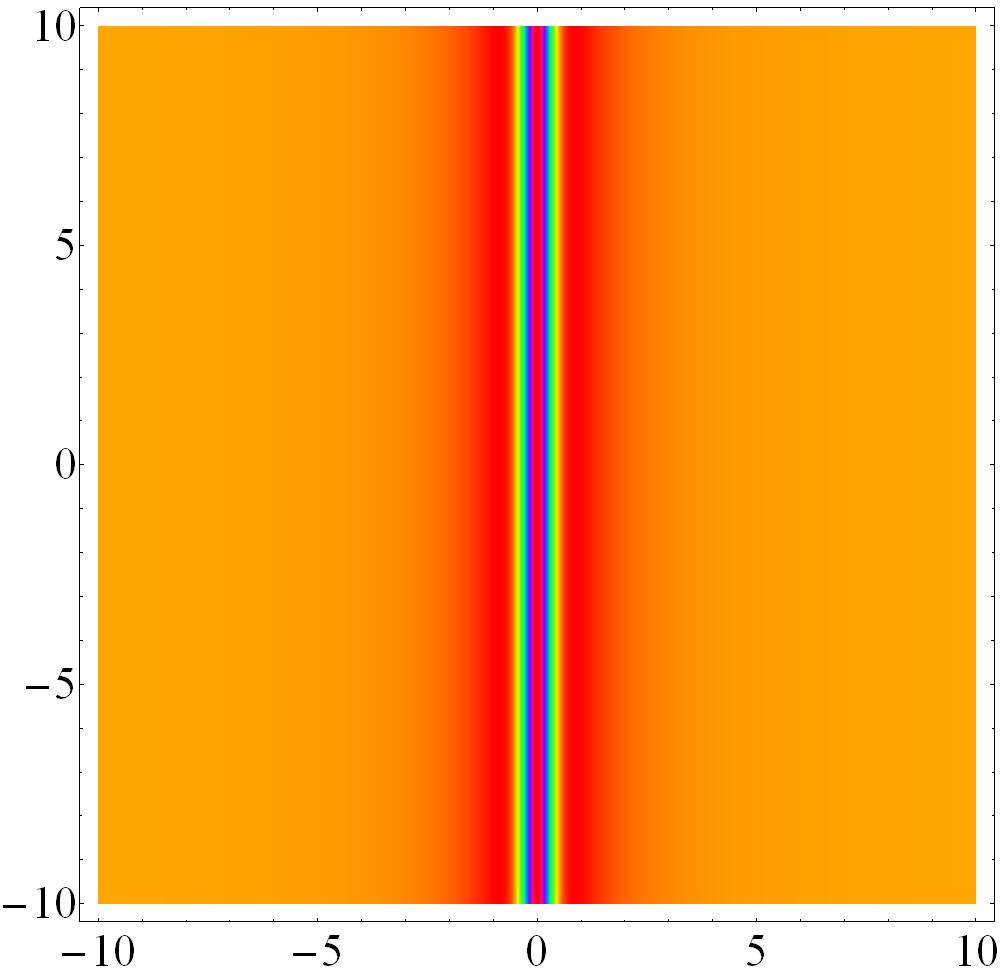}
\includegraphics[height=90 bp]{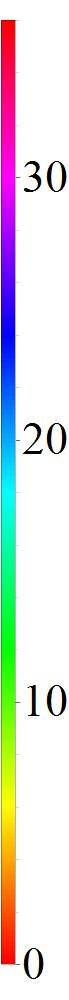}}
\quad
\subfigure[]{\includegraphics[width=130 bp]{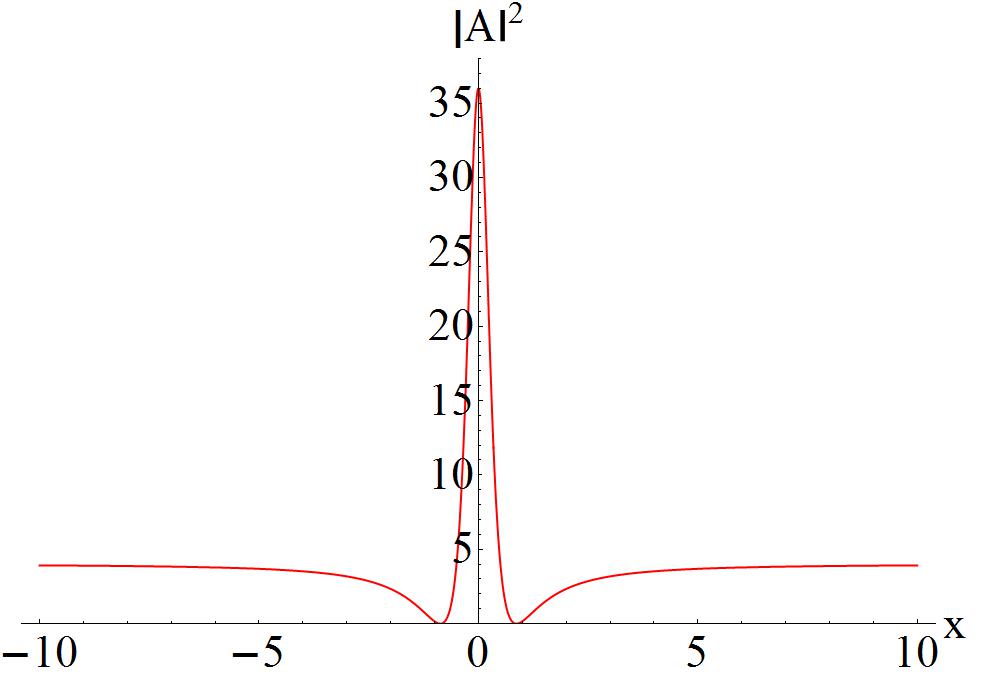}}
\caption{\footnotesize  A rogue wave transformed into a W-shaped soliton with
\,$a=2,\, b=1,\, \gamma=1,\,\omega=1,\,\alpha=-1,\,\beta=1,\,k_{1}=1,\,k_{2}=1\,$ and $\, \lambda_{1}=\lambda_{2}^{*}=-0.5+\,i.$ (b) is the contour plot of (a). (c) is the  cross-sectional view  of (a) at $t=0$. }
\end{figure}

In particular, as the period  $h_{R}$ is close to zero, namely, $n\rightarrow a\sqrt{\beta\gamma}$, the periodic
wave will turn into the W-shaped soliton, as
shown in \textbf{Fig.~7}. In this case, the solution~(\ref{EXP-B1}) is  transformed into
\begin{equation}\label{RWSA}
\qquad\qquad\qquad A_{RS}^{[1]}=-\frac{a \,e^{i x \omega } \left(a^2\, \beta \, \gamma  \,x^2-3\right)}{a^2 \,\beta \, \gamma\,  x^2+1}\,.
\end{equation}
The maximum height ($9a^{2}$) of the W-shaped wave is nine times the
background intensity while the minimum is zero. By comparison with the W-shaped soliton in Fig.~2,  the soliton in Fig.~7 has two different features: (1) it shows a single-peak structure without oscillating tails; (2) it has a  rational expression.

We further discuss the effects of the coefficients $\alpha$, $\beta$ and $\gamma$ on the transformed solitons. The coefficients $\alpha$, $\beta$ and $\gamma$, which are related to the parameters $c_{1}$, $c_{2}$, $l_{1}$ and $l_{2}$,  are defined by the equation~(\ref{AB11}).
\textbf{Figs.~(8a)} and \textbf{(8b)} indicate that the amplitudes of
the multi-peak solitons decrease with increasing the values of $\beta$ and $\gamma$. In addition, increasing the  values of $\beta$ and $\gamma$ leads to  stronger localization and a smaller oscillation period for the multi-peak soliton. This means that we can control the amplitudes of the transformed nonlinear waves by adjusting the  two group velocity coefficients $c_{1}$ and $c_{2}$, and the parameter $l_{2}$ that reflects the interaction of the wave
packet and the meanflow. However, as shown in \textbf{Fig.~(8c)}, changing the value of $\alpha$ doesn't affect the characteristics of the multi-peak soliton.
\begin{figure}[H]
\centering
\subfigure[]{\includegraphics[width=150 bp]{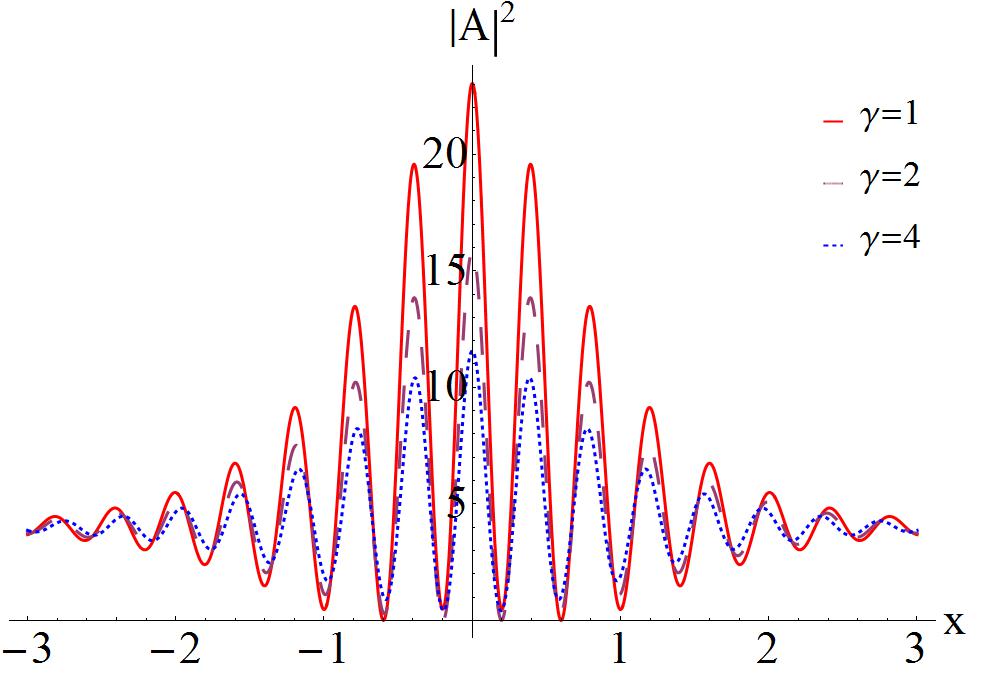}}
\quad
\subfigure[]{\includegraphics[width=150 bp]{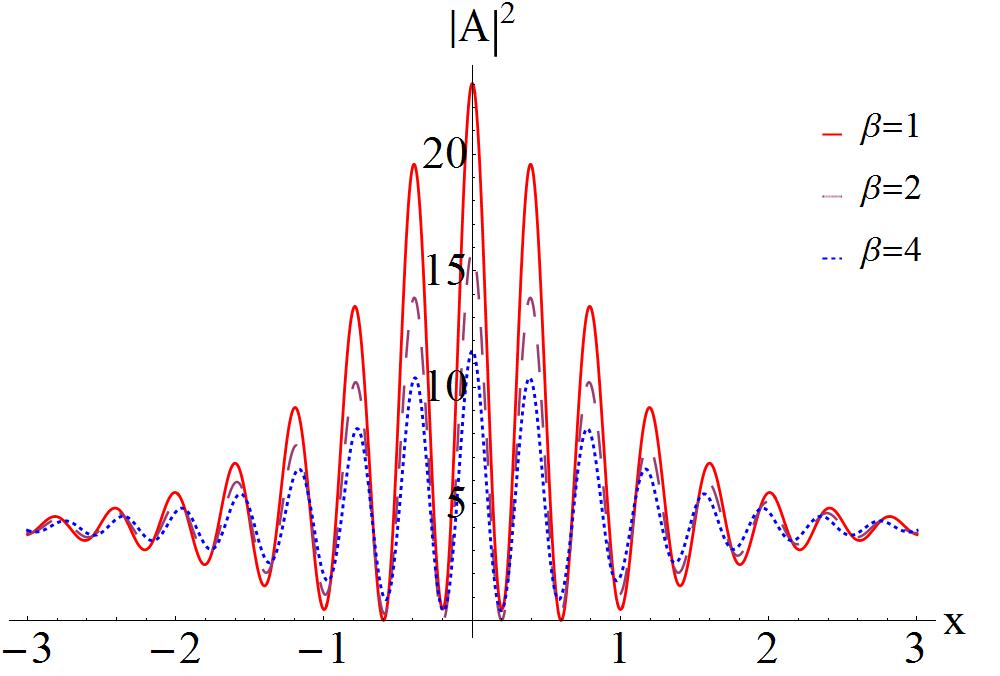}}
\quad
\subfigure[]{\includegraphics[width=150 bp]{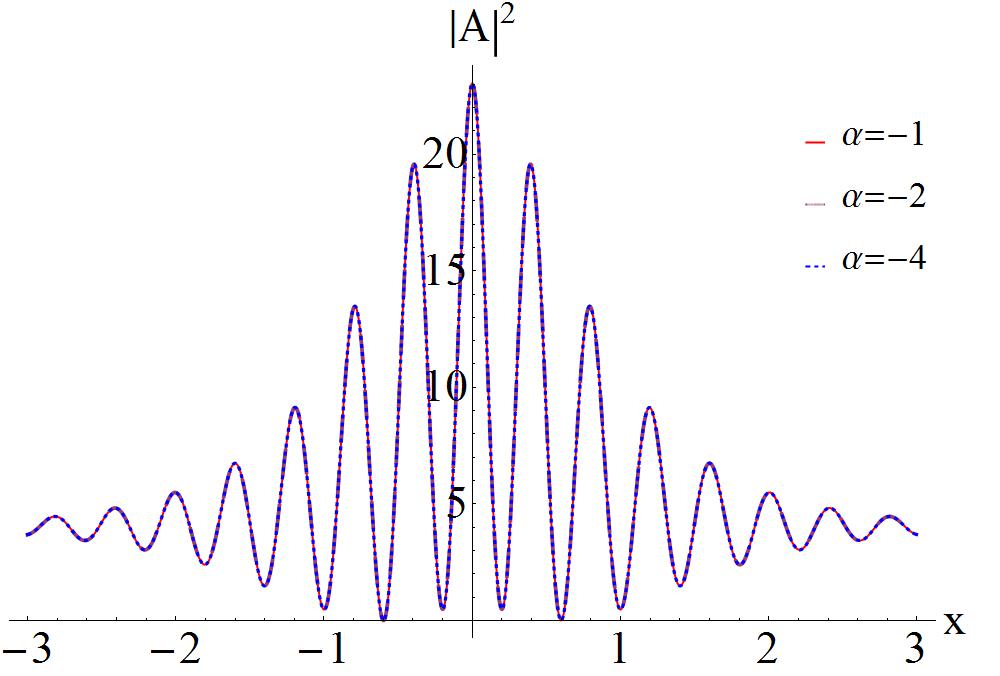}}
\caption{\footnotesize The effects of $\alpha$, $\beta$ and $\gamma$ on the multi-peak solitons with
\,$a=2,\, \omega=1,\,k_{1}=1,\,k_{2}=-1\,$ and $\, \lambda_{1}=\lambda_{2}^{*}=7.2-0.7\,i.\,(a): \alpha=-1,\,\beta=1; (b): \alpha=-1,\,\gamma=1; (c): \beta=1,\,\gamma=1.$
}
\end{figure}

\vspace{5mm}
\noindent\textbf{\Large{\uppercase\expandafter{3}. Nonlinear wave interactions}}

In this section, we study the characteristics of interactions between the nonlinear waves presented above. Due to the diversity of the transformed waves (in fact,  there are many  kinds of nonlinear  wave interactions), we only exhibit several typical nonlinear superposition patterns that derive from  the two-breather solutions.

By virtue of the $n$-fold DT~\cite{GR,WL}, the two-breather solution of System~(\ref{AB}) is given as
\begin{equation}\label{two-oreder}
\begin{aligned}
 A^{[2]}_{B}=A^{[0]}+\frac{4\,i}{\sqrt{\beta\gamma}}\,\frac{\Delta_{1}^{[2]}}{\Delta^{[2]}}\,, \qquad
&B^{[2]}_{B}=B^{[0]}-\frac{4\,i}{\beta}\,\Big(\frac{\Delta_{2}^{[2]}}{\Delta^{[2]}}\Big)_{t}\,,
\end{aligned}
\end{equation}
with
 {\begin{subequations}\begin{gather}
\quad A^{[0]}=a\,e^{i\,\rho}\,,\quad B^{[0]}=b\,,\nonumber\\
\quad \lambda_{1}=\lambda_{2}^{*}=m_{1}+n_{1}\,i,\quad
\lambda_{3}=\lambda_{4}^{*}=m_{2}+n_{2}\,i, \nonumber\\
\quad \psi_{2}=-\varphi_{1}^{*},\quad
\varphi_{2}=\psi_{1}^{*};\quad
\psi_{4}=-\varphi_{3}^{*},\quad
\varphi_{4}=\psi_{3}^{*};\nonumber \\
\quad
\varphi_{j}={k_{1}}\frac{i\,h_{j}+2 \,i\, \lambda_{j}+i\,\omega}{a \sqrt{\beta  \gamma}}\,e^{i\,(\varpi_{j}+\frac{\rho}{2})}+k_{2}\,e^{-i(\varpi_{j}-
\frac{\rho}{2})}\,,\nonumber\\
\quad \psi_{j}=k_{1}\,e^{i\,(\varpi_{j}-\frac{\rho}{2})}+k_{2}\,\frac{i\,h_{j}+2 \, i \,\lambda_{j}+i \,\omega }{a
   \sqrt{\beta  \gamma }}\,e^{-i(\varpi_{j}+\frac{\rho}{2})}\,,\nonumber\\ \quad
   j=1,3,
\quad k_{1}=k_{2}=1, \nonumber \\
\quad \Delta_{1}^{[2]}=
\begin{vmatrix}{}
\lambda_{1}\varphi_{1}&\varphi_{1}&-\lambda_{1}^{2}\varphi_{1}&\psi_{1} \nonumber\\
\lambda_{2}\varphi_{2}&\varphi_{2}&-\lambda_{2}^{2}\varphi_{2}&\psi_{2}\nonumber \\
\lambda_{3}\varphi_{3}&\varphi_{3}&-\lambda_{3}^{2}\varphi_{3}&\psi_{3}\nonumber \\
\lambda_{4}\varphi_{4}&\varphi_{4}&-\lambda_{4}^{2}\varphi_{4}&\psi_{4}
\end{vmatrix}\,,
\quad
\Delta_{2}^{[2]}=
\begin{vmatrix}{}
-\lambda_{1}^{2}\varphi_{1}&\varphi_{1}&\lambda_{1}\psi_{1}&\psi_{1}\nonumber \\
-\lambda_{2}^{2}\varphi_{2}&\varphi_{2}&\lambda_{2}\psi_{2}&\psi_{2}\nonumber \\
-\lambda_{3}^{2}\varphi_{3}&\varphi_{3}&\lambda_{3}\psi_{3}&\psi_{3}\nonumber \\
-\lambda_{4}^{2}\varphi_{4}&\varphi_{4}&\lambda_{4}\psi_{4}&\psi_{4}
\end{vmatrix}\,,\\\quad
\Delta^{[2]}=
\begin{vmatrix}{}
\lambda_{1}\varphi_{1}&\varphi_{1}&\lambda_{1}\psi_{1}&\psi_{1} \nonumber\\
\lambda_{2}\varphi_{2}&\varphi_{2}&\lambda_{2}\psi_{2}&\psi_{2}\nonumber \\
\lambda_{3}\varphi_{3}&\varphi_{3}&\lambda_{3}\psi_{3}&\psi_{3}\nonumber \\
\lambda_{4}\varphi_{4}&\varphi_{4}&\lambda_{4}\psi_{4}&\psi_{4}
\end{vmatrix}\,.
\end{gather}\end{subequations}}
\textbf{Fig.~9} displays the elastic interaction between two breathers of System~(\ref{AB}). It is found that the two-breather interaction produces a second-order central rogue wave in ($x$-$t$) plane.
Similar to the case in Section 2, to convert the two-breather solutions into the two-soliton ones on constant backgrounds, the parameter $b$ also needs to meet the transition condition~(\ref{KZFC}), namely, $k=0$. For different kinds of nonlinear superpositions, we can choose the corresponding real ($m_{j}$) and imaginary ($n_{j}$) parts of eigenvalues ($\lambda_{j}$)  in the solution~(\ref{two-oreder}). The values of $m_{j}$ and $n_{j}$ ($j=1, 3$)
control the types of the transformed  nonlinear waves. In the following, we present six kinds of nonlinear superposition patterns.

\begin{figure}[H]
\centering
\includegraphics[width=180 bp]{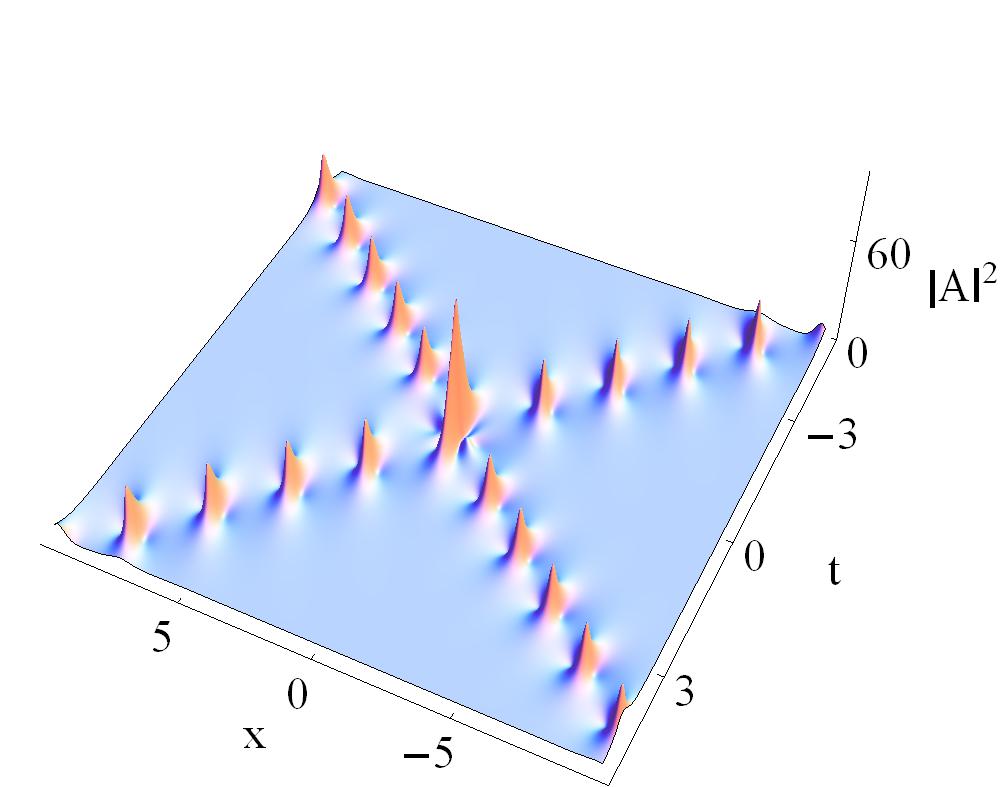}
\caption{\footnotesize  Elastic collision between two breathers with
\,$a=2,\, b=1,\, \gamma=4,\,\omega=1,\,\alpha=4,\,\beta=1,\,k_{1}=1,\,k_{2}=1\,,\, \lambda_{1}=\lambda_{2}^{*}=1.8\,i$ and $\, \lambda_{3}=\lambda_{4}^{*}=-0.8+1.8\,i.$
}
\end{figure}

\textbf{The oscillation W-shaped soliton.}
We first consider the nonlinear superposition of two stationary W-shaped solitons,  which is exhibited in \textbf{Fig.~10}. The two eigenvalues $\lambda_{1}=-0.5+2.1\,i$ and $\lambda_{3}=-0.5+2.2\,i$ meet the conditions of the W-shaped solitons.
Interestingly, instead of a higher-order W-shaped soliton, these two localized waves form an oscillation W-shaped soliton with higher intensity ($|A(0, 0)|^{2}=4\,(1+n_{1}+n_{2})^{2}$). $|A|^{2}\rightarrow a^{2}$ by assuming $z \rightarrow \infty$, $t \rightarrow \infty $ which gives
the asymptotic plane. Due to the nonpropagating characteristic of the nonlinear waves, the nonlinear superposition
could show some novel features. The interaction between two same type waves generates a different one.

\begin{figure}[H]
\centering
\subfigure[]{\includegraphics[width=180 bp]{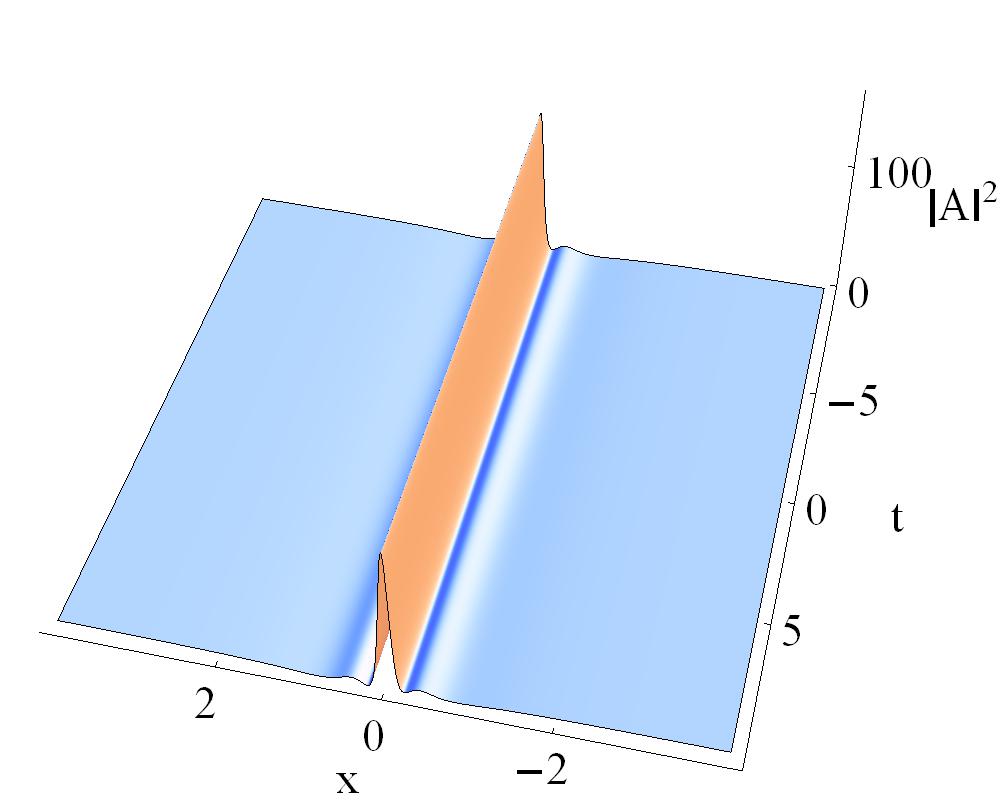}}
\quad
\subfigure[]{\includegraphics[width=150 bp]{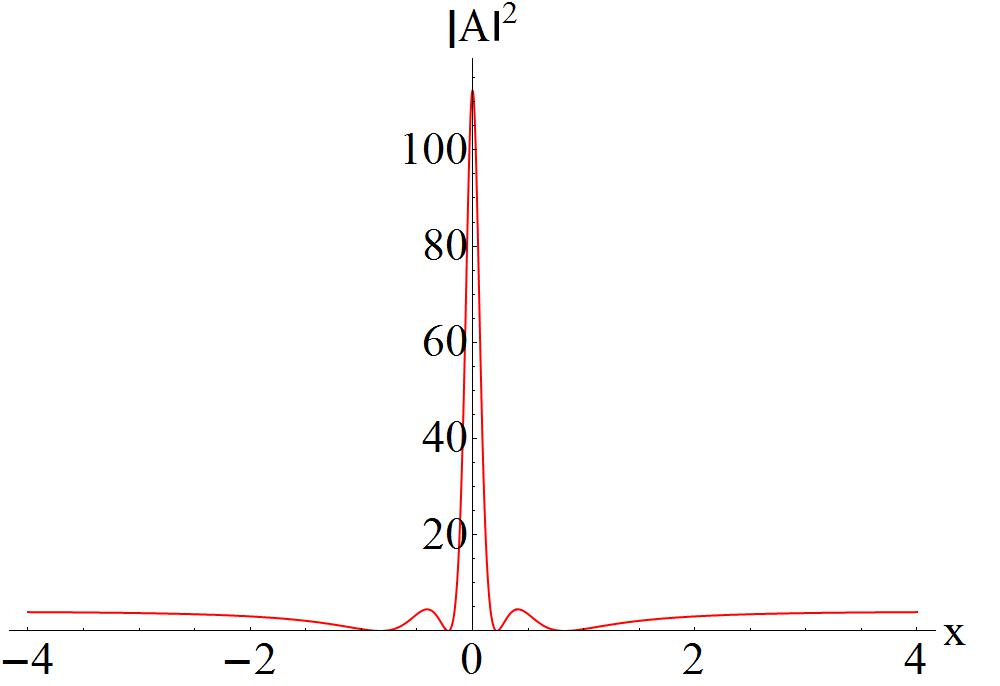}}
\caption{\footnotesize  The interaction between two W-shaped solitons  with
\,$a=2,\, b=1,\, \gamma=4,\,\omega=1,\,\alpha=-1,\,\beta=1,\,k_{1}=1,\,k_{2}=1\,,\, \lambda_{1}=\lambda_{2}^{*}=-0.5+2.1\,i$ and $\, \lambda_{3}=\lambda_{4}^{*}=-0.5+2.2\,i.$ The nonlinear interaction forms an oscillation W-shaped soliton.
}
\end{figure}

\textbf{The W-shaped soliton.} Next, we demonstrate the nonlinear interaction between two antidark solitons in \textbf{Fig.~11}. The two eigenvalues $\lambda_{1}=m_{1}+i\,n_{1}$ and $\lambda_{3}==m_{3}+i\,n_{3}$ are set to be $-0.5-2.2\,i$ and $-0.5-2.3\,i$, respectively.  Surprisingly, we find that the interaction between the antidark solitons doesn't produce a second-order antidark soliton, but instead  generates a W-shaped soliton.

\begin{figure}[H]
\centering
\subfigure[]{\includegraphics[width=180 bp]{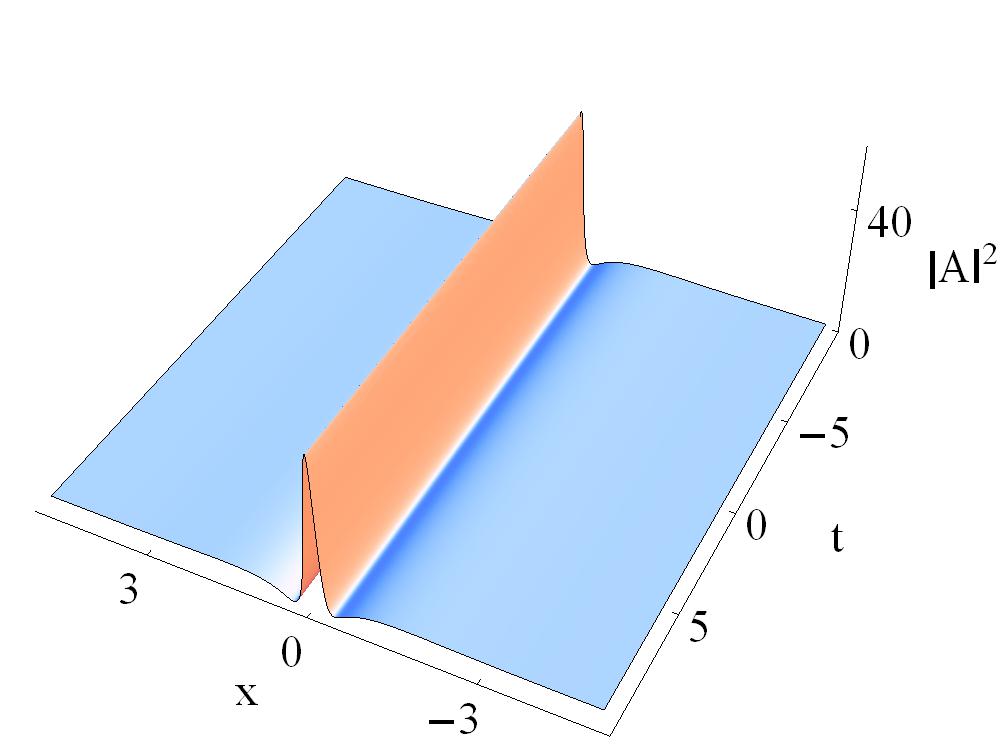}}
\quad
\subfigure[]{\includegraphics[width=150 bp]{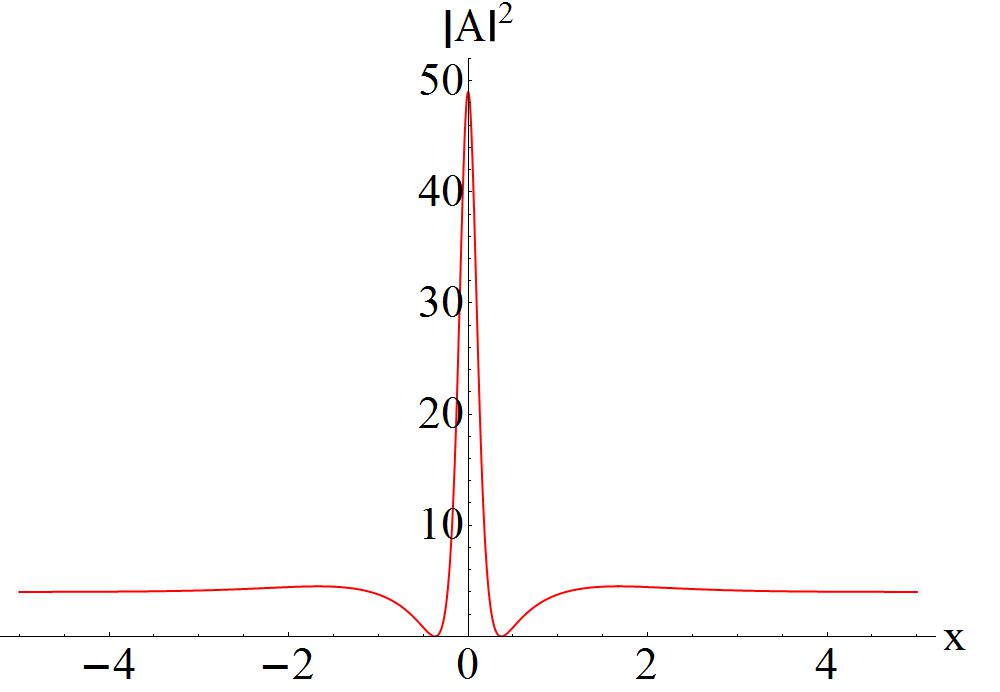}}
\caption{\footnotesize The interaction between two antidark solitons with
\,$a=2,\, b=1,\, \gamma=4,\,\omega=1,\,\alpha=-1,\,\beta=1,\,k_{1}=1,\,k_{2}=1\,,\, \lambda_{1}=\lambda_{2}^{*}=-0.5-2.2\,i$ and $\, \lambda_{3}=\lambda_{4}^{*}=-0.5-2.3\,i.$ The nonlinear interaction forms a W-shaped soliton.
}
\end{figure}

\textbf{The antidark soliton and W-shaped soliton pair.} We then study the interaction between  two different types of nonlinear waves, for example, the antidark soliton and a W-shaped soliton. The eigenvalue $\lambda_{1}=-0.5-2.1\,i$ is for the antidark soliton while  $\lambda_{3}=-0.5+4\,i$ for the W-shaped one. As shown in \textbf{Fig.~12}, an intriguing phenomenon is that the interaction forms an antidark soliton. By adjusting the values of eigenvalues $\lambda_{1}$ (still for antidark soliton) and $\lambda_{3}$ (still for W-shaped soliton), we observe a W-shaped soliton pair in \textbf{Fig.~13}. Each component in the W-shaped soliton pair has the same intensity. In spite of the similar single wave as the seeds,  the nonlinear interactions can produce different patterns due to the choices of different eigenvalues $\lambda_{1}$ and $\lambda_{3}$. This  means  the mechanism of nonlinear superposition is controlled by the two eigenvalues. A complete and analytical study of the control mechanism of such composite solutions is an issue that requires special investigation.

\begin{figure}[H]
\centering
\subfigure[]{\includegraphics[width=180 bp]{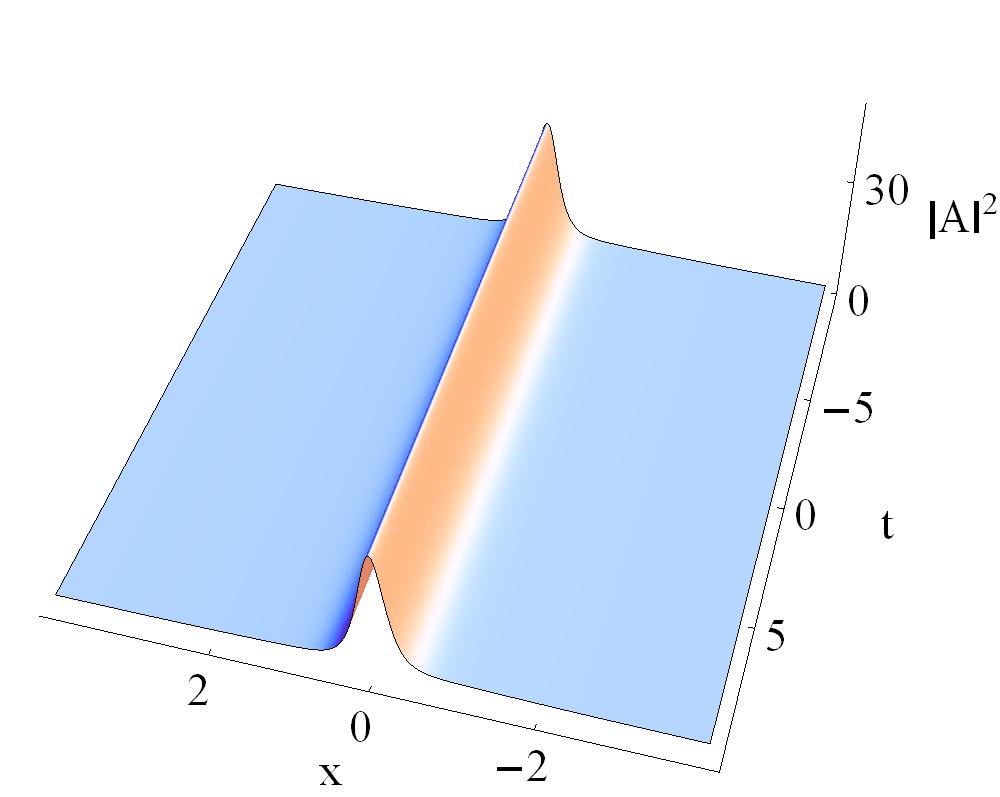}}
\quad
\subfigure[]{\includegraphics[width=150 bp]{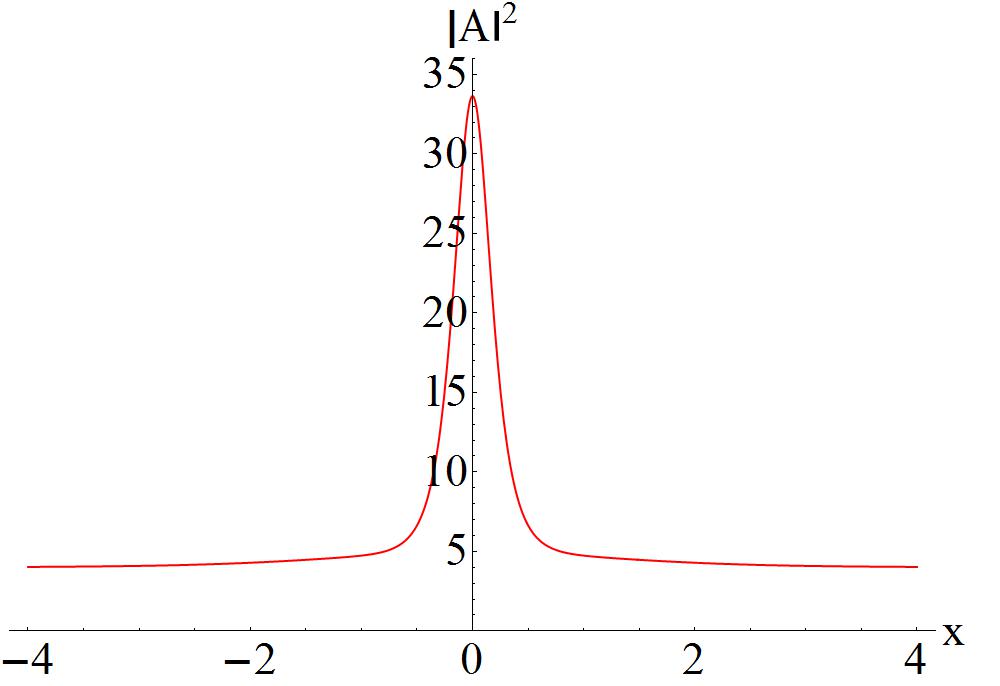}}
\caption{\footnotesize The interaction between an antidark soliton and a W-shaped soliton with
\,$a=2,\, b=1,\, \gamma=4,\,\omega=1,\,\alpha=-1,\,\beta=1,\,k_{1}=1,\,k_{2}=1\,,\, \lambda_{1}=\lambda_{2}^{*}=-0.5-2.1\,i$ and $\, \lambda_{3}=\lambda_{4}^{*}=-0.5+4\,i.$
}
\end{figure}

\begin{figure}[H]
\centering
\subfigure[]{\includegraphics[width=180 bp]{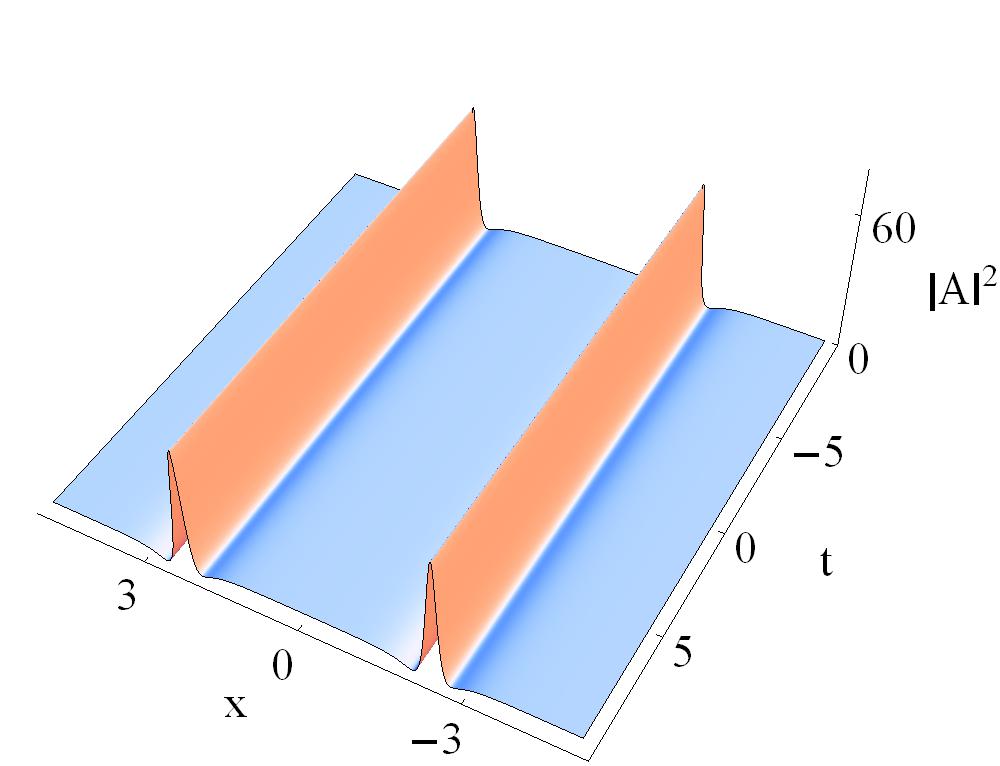}}
\quad
\subfigure[]{\includegraphics[width=150 bp]{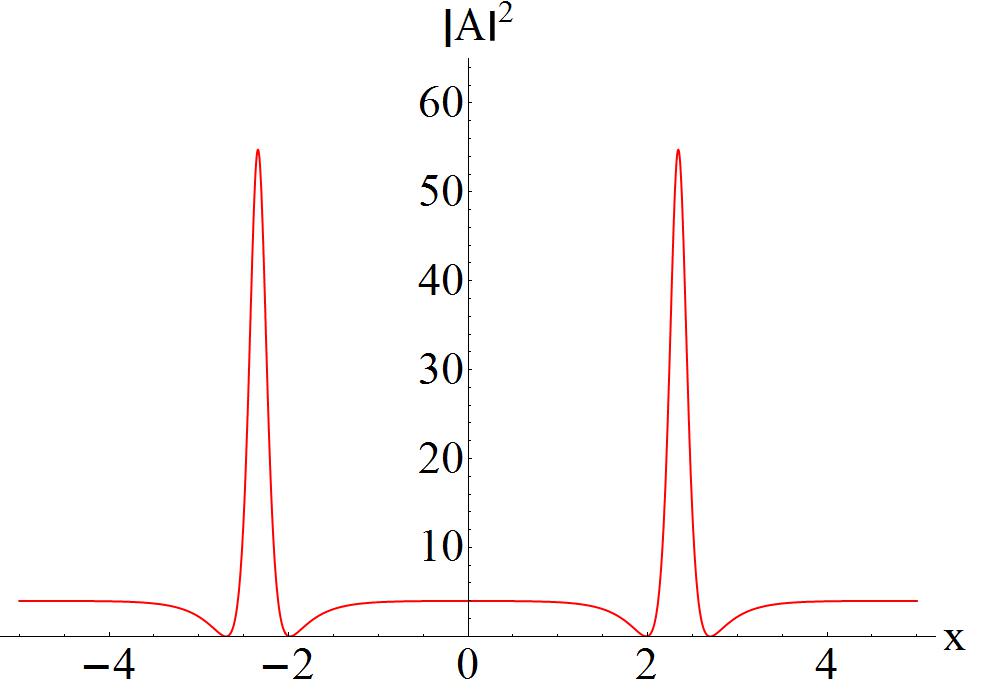}}
\caption{\footnotesize The interaction between  an antidark soliton and a W-shaped soliton with
\,$a=2,\, b=1,\, \gamma=4,\,\omega=1,\,\alpha=-1,\,\beta=1,\,k_{1}=1,\,k_{2}=1\,,\, \lambda_{1}=\lambda_{2}^{*}=-0.5-2.7\,i$ and $\, \lambda_{3}=\lambda_{4}^{*}=-0.5+2.699\,i.$
}
\end{figure}

\textbf{The multi-peak solitions.} Finally, we display some multi-peak structures.
\textbf{Fig.~14} describes the nonlinear interactions between the oscillation W-shaped soliton and  oscillation M-shaped soliton. The  superposition leads to the formation of a three-peak soliton that corresponds to the case $|A(0, 0)|^{2}_{xx}>0$. Using different eigenvalues, we plot a four-peak soliton ($|A(0, 0)|^{2}_{xx}<0$) also produced by the  oscillation W-shaped soliton and  oscillation M-shaped soliton, as depicted in \textbf{Fig.~15}. A higher-order multi-peak soiton, which is composed of two first-order multi-peak solitons, is plotted in \textbf{Fig.~16}.

\begin{figure}[H]
\centering
\subfigure[]{\includegraphics[width=180 bp]{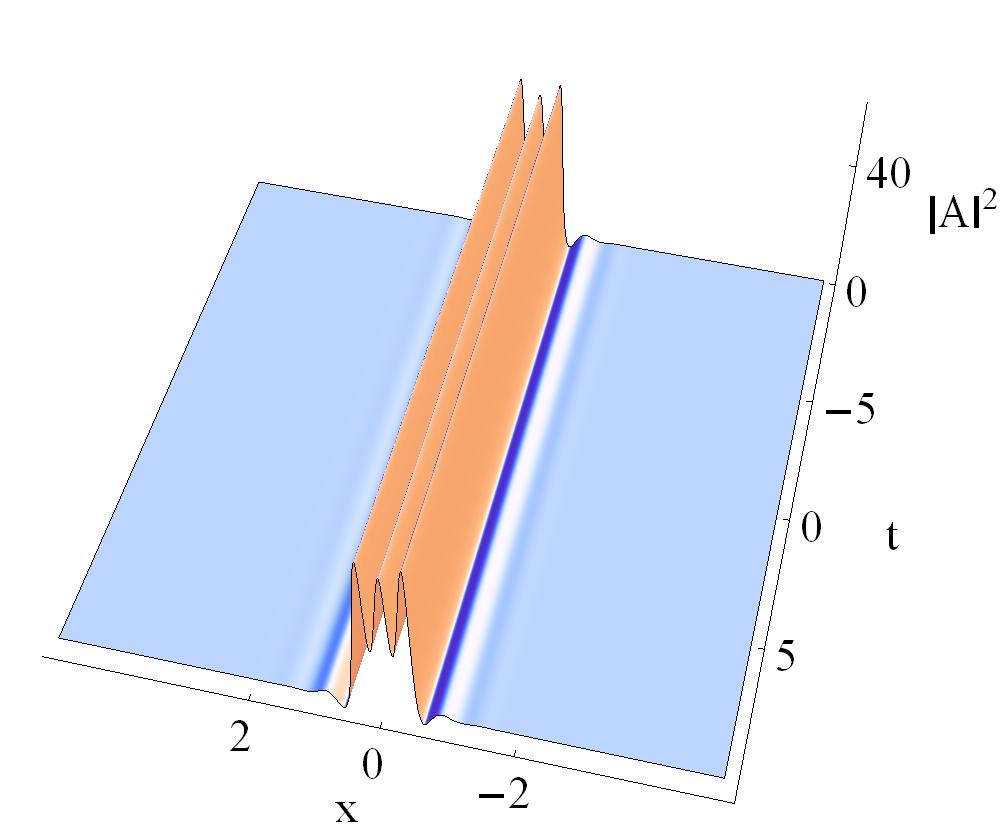}}
\quad
\subfigure[]{\includegraphics[width=150 bp]{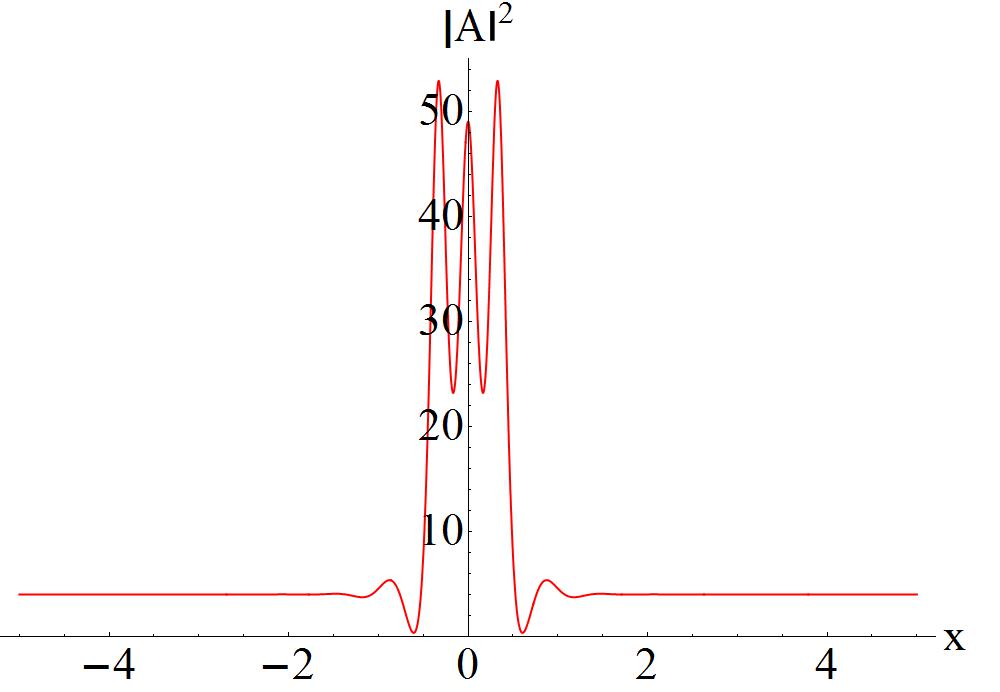}}
\caption{\footnotesize The interaction between an oscillation W-shaped soliton and an oscillation M-shaped soliton  with
\,$a=2,\, b=1,\, \gamma=4,\,\omega=1,\,\alpha=-1,\,\beta=1,\,k_{1}=1,\,k_{2}=1\,,\, \lambda_{1}=\lambda_{2}^{*}=4.5-2.7\,i$ and $\, \lambda_{3}=\lambda_{4}^{*}=5.5+5.2\,i.$
}
\end{figure}

\begin{figure}[H]
\centering
\subfigure[]{\includegraphics[width=180 bp]{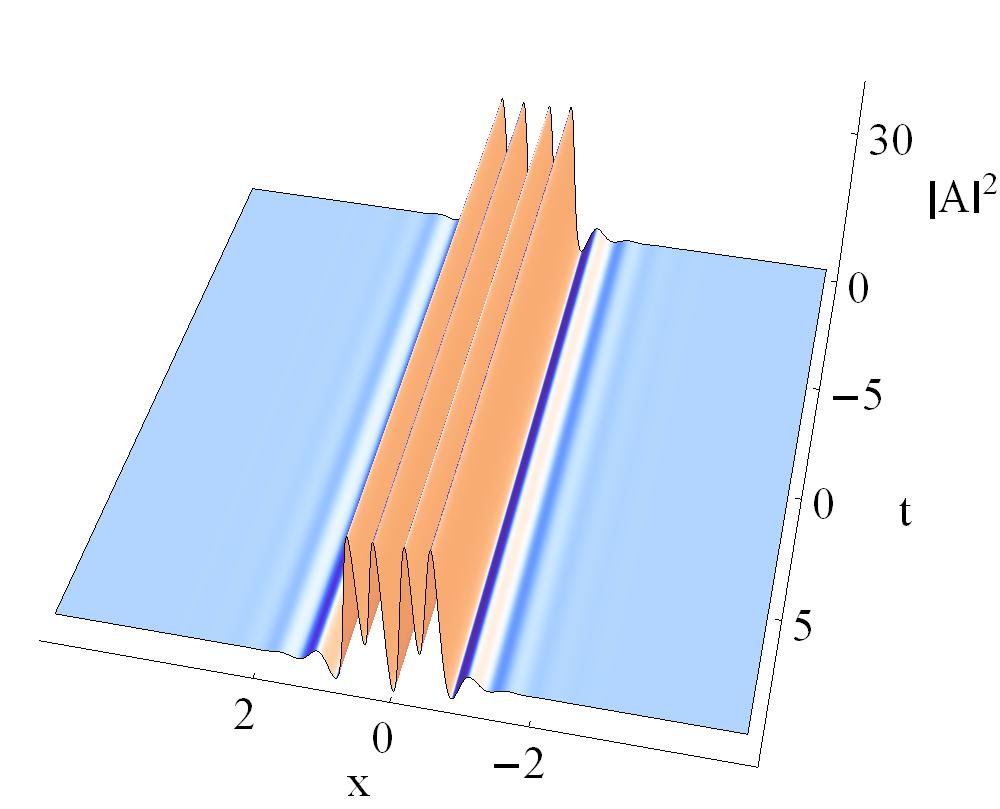}}
\quad
\subfigure[]{\includegraphics[width=150 bp]{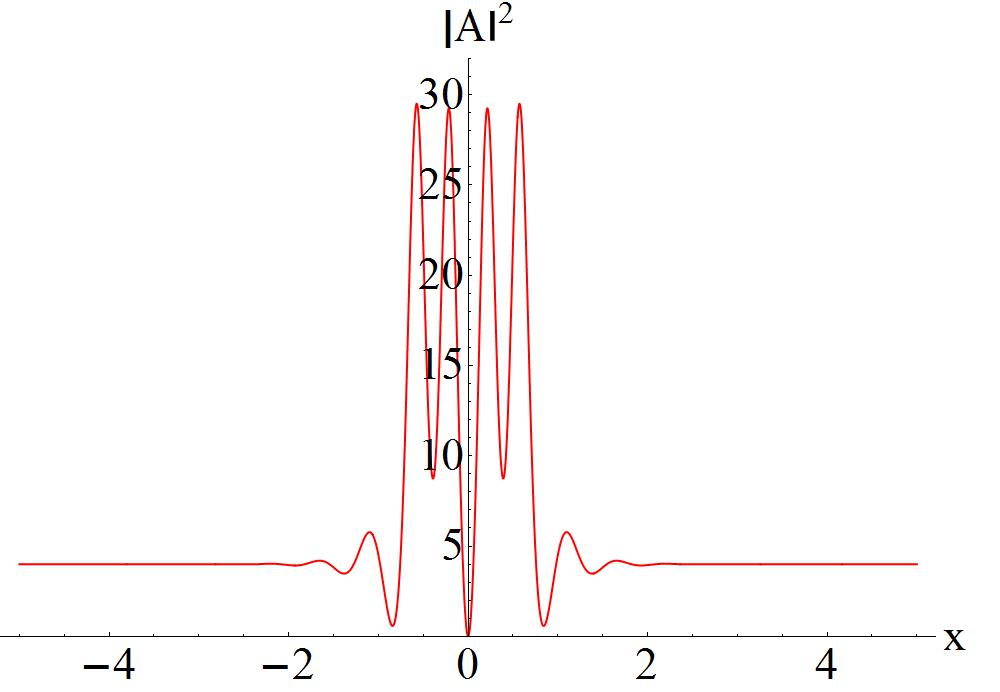}}
\caption{\footnotesize The interaction between an oscillation W-shaped soliton and an oscillation M-shaped soliton with
\,$a=2,\, b=1,\, \gamma=4,\,\omega=1,\,\alpha=-1,\,\beta=1,\,k_{1}=1,\,k_{2}=1\,,\, \lambda_{1}=\lambda_{2}^{*}=6-3.05\,i$ and $\, \lambda_{3}=\lambda_{4}^{*}=5+2\,i.$
}
\end{figure}

\begin{figure}[H]
\centering
\subfigure[]{\includegraphics[width=180 bp]{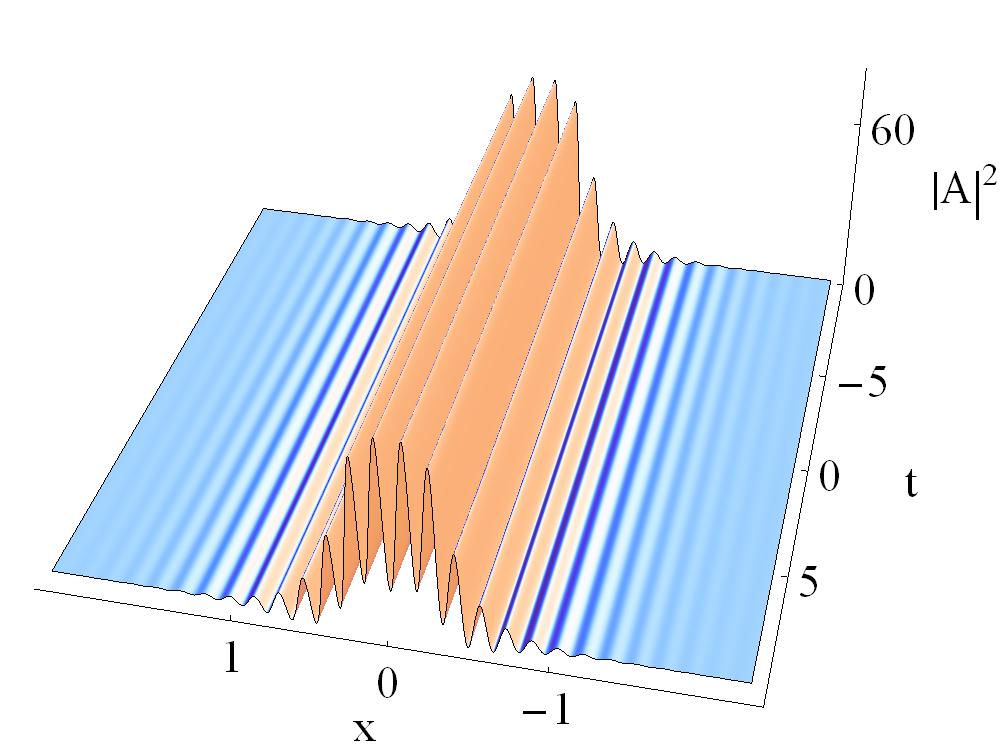}}
\quad
\subfigure[]{\includegraphics[width=150 bp]{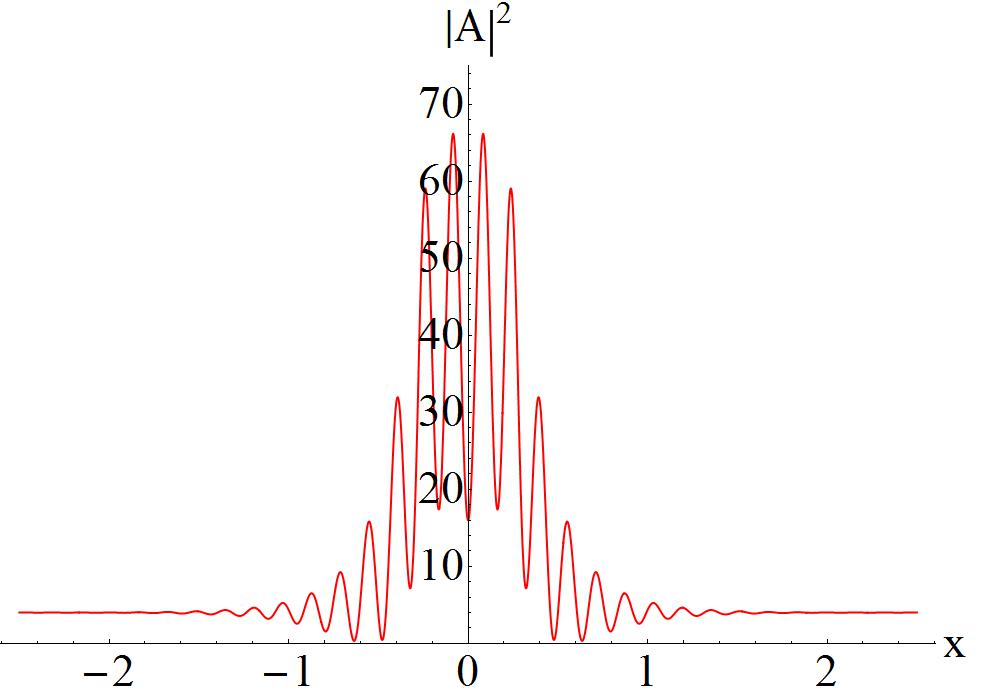}}
\caption{\footnotesize The interaction between two multi-peak solitons with
\,$a=2,\, b=1,\, \gamma=4,\,\omega=1,\,\alpha=-1,\,\beta=1,\,k_{1}=1,\,k_{2}=1\,,\, \lambda_{1}=\lambda_{2}^{*}=-19-5\,i$ and $\, \lambda_{3}=\lambda_{4}^{*}=-20+2\,i.$
}
\end{figure}

\vspace{5mm}
\noindent\textbf{\Large{\uppercase\expandafter{4}. MI characteristics}}

In this section, we reveal the explicit relation between the transition and the distribution characteristics of  MI growth rate for System~(\ref{AB}). Firstly,
let us recall the results of MI analysis on  System~(\ref{AB}). As shown in Ref.~\cite{WL}, System~(\ref{AB}) admits the following continuous wave solutions,
\begin{equation}\label{EXP-MI1}
\begin{aligned}
\qquad\qquad\qquad\qquad\quad\quad\quad&{A}(x,t)=a\,e^{i\,(\omega\,x+k\,t)}\,,\\
\qquad\qquad\qquad\qquad\quad&{B}(x,t)=b\,,
\end{aligned}
\end{equation}
where $ a $,\,$ \omega $,\,$ k $ and $ b $ are real parameters. A perturbed nonlinear background can be expressed as
\begin{equation}\label{EXP-MI2}
\begin{aligned}
\qquad\qquad\qquad\qquad\quad\quad&{A}(x,t)=(a+\epsilon\,{\widehat{A}}(x,t))\,e^{i\,(\omega\,x+k\,t)},\\
\qquad\qquad\qquad\qquad\quad\quad&{B}(x,t)=b+\epsilon\,{\widehat{B}}(x,t).
\end{aligned}
\end{equation}
Taking Eq.~(\ref{EXP-MI2}) into System~(\ref{AB}) yields the evolution equation for the perturbations as
\begin{equation}
\label{lse}
\begin{aligned}
\qquad\qquad&-a\,\beta\,{\widehat{B}}(x,t)+i\,k\,{\widehat{A}}^{(1,0)}(x,t)+i\,\omega\,{\widehat{A}}^{(0,1)}(x,t)+{\widehat{A}}^{(1,1)}(x,t)=0,\\
&{\widehat{B}}^{(1,0)}(x,t)+\frac{1}{2}\,a\,\gamma\,{\widehat{A}}^{(0,1)}(x,t)+\frac{1}{2}\,a\,\gamma\,{\widehat{A}}^{*(0,1)}(x,t)=0.
\end{aligned}
\end{equation}
Noting the linearity of Eq.~(\ref{lse})  with respect to $\widehat{A}$ and $\widehat{B}$, we introduce
\begin{equation}
\label{lse1}
\begin{aligned}
\qquad\qquad\qquad\qquad&{\widehat{A}}(x,t)=U\,\cos(\Lambda\,x-\Omega\,t)+i\,V\,\sin(\Lambda\,x-\Omega\,t)\,,\\
\qquad\qquad\qquad\qquad&{\widehat{B}}(x,t)=K\,\cos(\Lambda\,x-\Omega\,t)\,,\\
\end{aligned}
\end{equation}
which is characterized by the wave number $\Lambda$ and frequency $\Omega$. Using Eq.~(\ref{lse1}) into Eq.~(\ref{lse}) gives a linear homogeneous system of equations for $U$  and $K$:
\begin{equation}\label{18}
\begin{aligned}
\qquad\qquad\qquad\qquad-i\,a\,U\,\Lambda\,k+i\,a\,\omega\,U\,\Omega+i\,a\,V\,\Lambda\,\Omega=0,
\end{aligned}
\end{equation}
\begin{equation}\label{19}
\begin{aligned}
\qquad\qquad\qquad\qquad-a\,V\,\Lambda\,k+a\,\omega\,V\,\Omega-\frac{a^{2}\,U\,\beta\,\gamma\,\Omega}{\Lambda}+a\,U\,\Lambda\,\Omega=0.
\end{aligned}
\end{equation}
From the determinant of the coefficient matrix of Eqs.~(\ref{18})$\sim$(\ref{19}), the dispersion relation for the linearized disturbance can be determined as
{\begin{equation}
\qquad\qquad\qquad\qquad\Omega ^2 \left(-a \beta  \gamma +\Lambda
   ^2-\omega ^2\right)-k^2 \Lambda ^2+2 k
   \Lambda  \omega  \Omega =0.
\end{equation}}
Solving the above equation, we have
\begin{equation}
\qquad\qquad\qquad\qquad\Omega =\frac{k \Lambda  \omega \pm
   \sqrt{k^2 \Lambda ^2 \left(\Lambda
   ^2-a \beta  \gamma \right)}}{a \beta
   \gamma -\Lambda ^2+\omega ^2}.
\end{equation}

\begin{figure}[H]\centering
 {\includegraphics[width=110 bp]{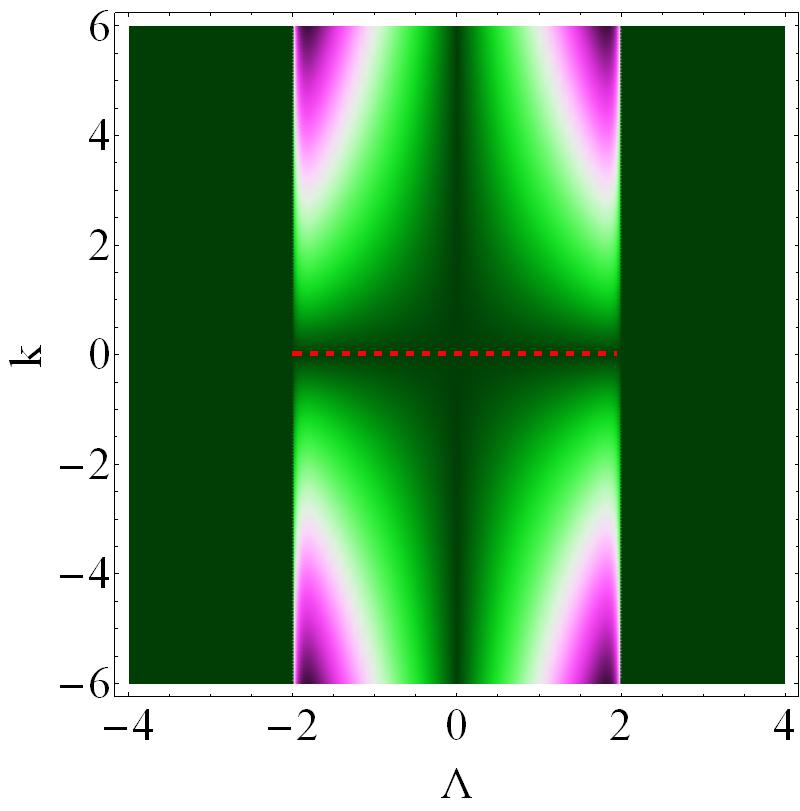}
  \includegraphics[height=110 bp]{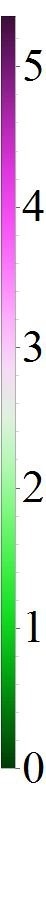}}

  \caption{\footnotesize Characteristics of MI growth rate $\Omega$ on $(\Lambda,k)$ plane with $a=1$, $\gamma=4$, $\beta=1$ and $\omega=1$. Here the dashed red lines represent  the stability region in
the perturbation frequency region $-a\,\sqrt{\beta\gamma}<\Lambda<a\,\sqrt{\beta\gamma} $ , which is given as\, $b=-\frac{\alpha}{\beta}$.}
\end{figure}

\textbf{Fig.~17} shows the characteristics of MI on the ($\Lambda, k$) plane. It is found that the MI exists in the region $-a\,\sqrt{\beta\gamma}<\Lambda<a\,\sqrt{\beta\gamma}$. Ref.~\cite{WL} has found that the rogue wave existence condition of System~(\ref{AB}) is strictly related to
baseband MI, namely, the MI whose bandwidth includes arbitrary small frequencies. In this case, we obtain the  parameter condition $\beta\gamma>0$~\cite{WL}.
Hereby, we discover that the MI growth rate distribution is symmetric with respect to
\begin{equation}\label{MIE}
k=0\,,
\end{equation}
i.e.,
\begin{equation}\label{MIE1}
b=-\frac{\alpha}{\beta}.
\end{equation}
The $k=0$ line (red dashed line) in Fig.~17  corresponds to a modulational stability (MS) region where
the growth rate is vanishing in the low perturbation frequency region.
More interestingly, one can find that the MS condition~(\ref{MIE}) [or (\ref{MIE1})] is completely consistent with the condition~(\ref{KZFC})
 which converts breathers into stationary nonlinear waves. Our finding suggests that the transition between breathers and stationary nonlinear waves can occur in the
MS region with the low frequency perturbations.

\vspace{5mm}
\noindent\textbf{\Large{\uppercase\expandafter{5}. Conclusions}}

In summary, we have investigated the transition between breathers and stationary nonlinear waves for System~(\ref{AB}). Some intriguing different types of stationary nonlinear waves, including multi-peak solitons, M-shaped soliton, periodic wave, antidark soliton and W-shaped soliton, have been demonstrated graphically and analytically. The reasons and the conditions for transition have been presented explicitly. We have found that the real part of eigenvalue $m$ has effects on the peak numbers of the multi-peak solitons while the imaginary part $n$ controls the transition between the  oscillating W-shaped soliton and M-shaped soliton. Further, we have studied the interactions between those transformed nonlinear waves. Due to the  nonpropagating characteristic, the
 nonlinear superposition of two types of nonlinear waves have exhibited some novel features (see Figs.~10$\sim$16). Finally, we have shown that this transition is strictly associated with the MI analysis that involves a MS region.

\vspace{5mm}
\noindent\textbf{\Large{ Acknowledgements}}

We express our sincere thanks to all the members of our discussion group for their valuable comments. This work has been supported by the National Natural Science Foundation of China under Grant (Nos.~11305060 and 61505054), by the Fundamental Research Funds of the Central Universities (Project No.~2015ZD16), by the Innovative Talents Scheme of North China Electric Power University, and by the higher-level item cultivation project of Beijing Wuzi University
(No. GJB20141001).

\bibliography{abanov-bibliography}

\end{document}